\documentclass[aps,prd,onecolumn,showpacs,superscriptaddress,nofootinbib,longbibliography,eqsecnum,notitlepage]{revtex4-1}

\usepackage[utf8]{inputenc}
\usepackage{hyperref}
\usepackage{xcolor}
\usepackage{graphicx}
\usepackage{amsmath,amssymb}
\usepackage{bm}
\usepackage{floatrow}
\usepackage{capt-of}
\usepackage[font=small,labelfont=bf,tableposition=top]{caption}
\usepackage{booktabs}
\usepackage{varwidth}
\usepackage[export]{adjustbox}
\newfloatcommand{capbtabbox}{table}[][\FBwidth]
\DeclareCaptionLabelFormat{andtable}{#1~#2  \&  \tablename~\thetable}

\begin{document}
\title{Lensing anomaly and oscillations in the primordial power spectrum}

\author{Guillem Dom{\`e}nech}
\email{domenech@thphys.uni-heidelberg.de}  
\affiliation{Institut f\"ur Theoretische Physik, Ruprecht-Karls-Universit\"at Heidelberg, \\
Philosophenweg 16, 69120 Heidelberg, Germany}
\author{Marc Kamionkowski}
\email{kamion@jhu.edu}  
\affiliation{Department of Physics \& Astronomy, Johns Hopkins
University, 3400 N.\ Charles St., Baltimore, MD 21218, USA}

\begin{abstract}
The latest analysis of the cosmic microwave background by the
Planck team finds more smoothing of the acoustic peaks in the
temperature power spectrum than predicted by $\Lambda$CDM. Here
we investigate whether this additional smoothing can be mimicked by
an oscillatory feature, generated during inflation, that is
similar to the acoustic peaks but out of phase. We consider
oscillations generated by oscillating modulations of the
background---e.g., due to heavy fields or modulated
potentials---and by sharp features. We show that it is difficult
to induce oscillations that are linear (or almost linear) in $k$
by oscillatory modulations of the background.   We find, however,
that a sharp bumpy feature in the sound speed of perturbations
is able to produce the desired oscillations. The scenario can be
tested by combining CMB and BAO data.
\end{abstract}



\maketitle

\section{Introduction} \label{sec:intro}

As cosmic microwave background (CMB) photons travel towards
us, their trajectories are deflected by the gravitational
potentials generated by the matter distribution. This weak
lensing of the CMB has an impact on the CMB
temperature power spectrum \cite{Seljak:1995ve,Lewis:2006fu}.
The lensing magnifies the angular
size of the primordial fluctuations in some places on the sky
and de-magnifies others.  The observed peak {structure}
in the temperature power spectrum, when measured over the entire
sky, are therefore blurred \cite{Hanson:2009kr}: the acoustic
peaks are reduced slightly, and the troughs between them filled in.

Interestingly, when the theoretical prediction for this
smoothing is compared with the Planck data, it is found that the
lensing smoothing is larger than expected by roughly $10\%$
\cite{Aghanim:2018eyx}. The so-called $A_L$
anomaly\footnote{$A_L$ parametrizes a rescaling of the lensing
power spectrum such that $A_L=1$ for $\Lambda$CDM
\cite{Calabrese:2008rt}.} is persistent and recently slightly
more statistically significant, with a value $A_L=1.149 \pm
0.072$ ($68\%$ confidence), that constitutes a $2\sigma$ tension with
$\Lambda$CDM cosmology \cite{Aghanim:2018eyx}.  Moreover, the
residuals between the signal and the theoretical prediction
yield an oscillatory pattern whose frequency is roughly linear
in the multipole number $\ell$ and similar in shape to the
acoustic peaks.

If the tension persists with higher statistical significance,
it might be explained by some new physics that mimicks the
smoothing effect of lensing. One
possibility discussed by the Planck collaboration
\cite{Aghanim:2018eyx} is that there might be a component of
cold dark matter isocurvature (CDI) perturbation with a blue tilt. Since the
acoustic peaks of the CDI will have the opposite phase, this will
effectively smooth out the photon acoustic peaks. A similar mechanism was also studied in Refs.~\cite{Munoz:2015fdv,Smith:2017ndr} where the isocurvature perturbations of dark matter and baryons compensate each other. However, these models are tightly constraint by their effects on the trispectrum \cite{Smith:2017ndr}. Another
possibility is that there are oscillations in the primordial
power spectrum which have the same frequency but opposite phase
with the acoustic peaks \cite{Aghanim:2018eyx}. However, in an
analysis where the oscillatory feature in the power spectrum has
a $k$ independent amplitude and a frequency linear in $k$, no
correlation between the amplitude of the oscillations and $A_L$
is found \cite{Akrami:2018odb}. Moreover, one reaches similar conclusions when reconstructing the power spectrum from raw data \cite{Hazra:2014jwa,Hazra:2018opk}. In particular, it was already pointed out in Ref.~\cite{Hazra:2014jwa} that there is a degeneracy between the effects of lensing and oscillating features in the primordial power spectrum. On the theoretical side,
though, it is not clear whether physical models that might
induce wiggles in the primordial power spectrum are required to
do so with a scale-independent amplitude, nor with a precisely
linear dependence on $k$.  Previous fits to the residuals of the
temperature power spectrum were pursued in
Refs.~\cite{Achucarro:2014msa,Chen:2014joa,Hazra:2014goa,Hazra:2016fkm,Hazra:2017joc} but a mimicking of
the lensing effect from an inflationary model was not studied. Note that the reconstruction of the primordial power spectrum taking into account lensing also yields an oscillating feature \cite{Hazra:2018opk}. However, it is not clear what kind of inflationary model would give rise to such specific wiggle \cite{Hazra:2018opk}.

Here we explore inflationary models that might give rise to
oscillatory features in the primordial power spectrum that might
account for the $A_L$ anomaly.
Oscillatory features generated during inflation usually have an oscillation frequency which has a logarithmic or linear dependence on $k$ (see Ref.~\cite{Chluba:2015bqa} for a review). On
one hand, the logarithmic dependence could be either because
there is an oscillating modulation in the Lagrangian that
depends linearly on the inflaton which is slowly rolling
\cite{Pahud:2008ae,Flauger:2009ab} or because an extra massive
field is oscillating around its minimum in which case it
oscillates with a constant frequency and linearly in the cosmic
time \cite{Chen:2011zf,Chen:2014cwa,Chen:2015lza}. Also, successive turns in the multi-field inflationary trajectory yield a logarithmic dependence in $k$ \cite{Gao:2013ota}. On the other
hand, a frequency linear in $k$ is typical from sharp
transitions, e.g. steps in the potential or sudden turns in the
field space, with a damped amplitude depending on the sharpness
of the transition
\cite{Starobinsky:1992ts,Kamionkowski:1999vp,Gong:2001he,Stewart:2001cd,Adams:2001vc,Kaloper:2003nv,Choe:2004zg,Dvorkin:2009ne,Achucarro:2010da,Adshead:2011jq,Shiu:2011qw,Cespedes:2012hu,Gao:2013ota,Bartolo:2013exa,Achucarro:2014msa}. Note that sharp transitions can be understood in terms of Bogoliubov transformations \cite{Kaloper:2003nv}, since the negative frequency mode has been excited by the turn or step. It is also interesting to note that both logarithmic and linear $k$ dependences of the frequency may be related to a trans-planckian modulation \cite{Easther:2005yr}. There
is another interesting  
case where the
frequency goes as a power-law of $k$
\cite{Huang:2016quc,Domenech:2018bnf}; this may occur, e.g.,
when there is a background oscillation with a time dependent
frequency. See Refs.~\cite{Chen:2011zf,Chen:2018cgg} for a
power-law dependence in $k$ in alternatives scenarios to
inflation.

This paper is organized as follows. In Sec.~\ref{sec:review} we
review the $A_L$ anomaly and the requirement for the oscillatory
patterns in the power spectrum to mimick the lensing
smoothing. In Sec.~\ref{sec:features}, we study which features
could potentially yield such oscillatory patterns and we
conclude in Sec.~\ref{sec:conclusions}.  
We discuss possible models in Ap{}p.~\ref{app:models} and present details of the calculations in Apps.~\ref{App:spectrum} and \ref{App:trispectrum}.

\section{\texorpdfstring{$A_L\,$}a anomaly}\label{sec:review}

\begin{figure}
\includegraphics[width=0.49\columnwidth]{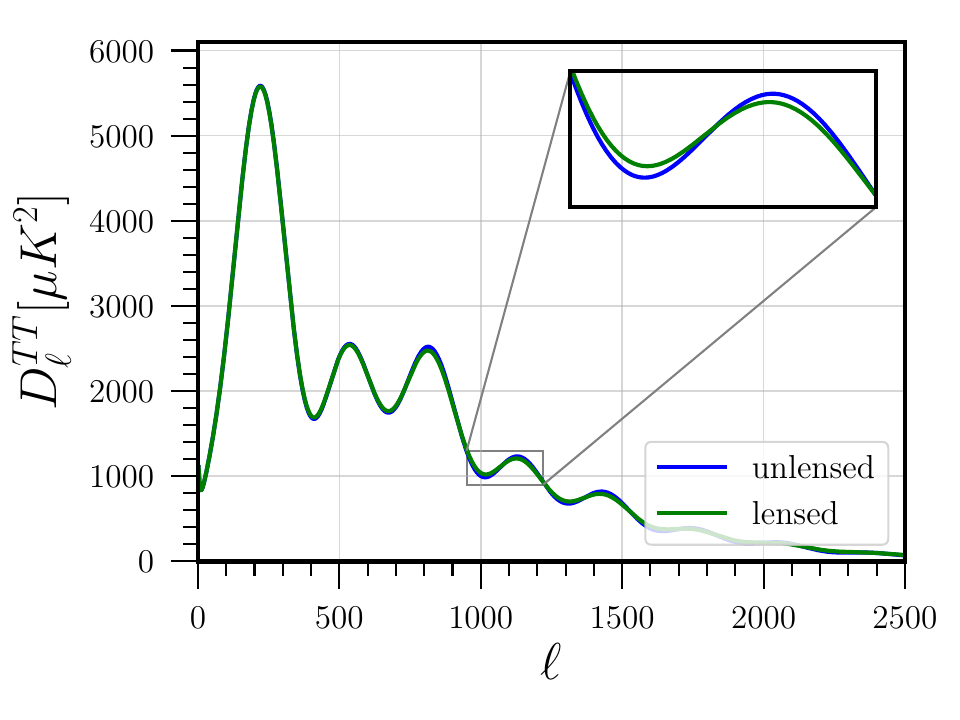}
\includegraphics[width=0.49\columnwidth]{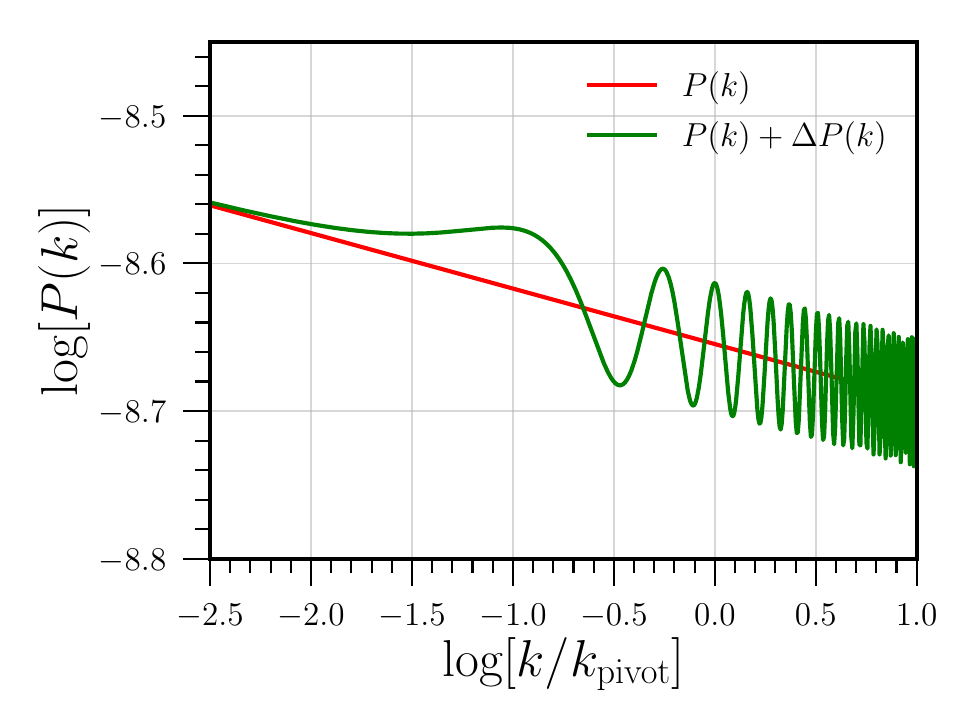}
\caption{On the left, CMB temperature power spectrum. We respectively show the unlensed and lensed power spectrum in blue and green. See how the acoustic peaks are smoothed out. On the right, the oscillatory feature linear in $k$ in the primordial power spectrum vs the usual power-law spectrum. \label{fig:lensed}}
\end{figure}

The effect of weak lensing onto the CMB power spectrum is to
smooth out the acoustic peaks by blurring the acoustic-peak
structure in $\ell$ space (see Fig.~\ref{fig:lensed}). Using the modelled unlensed CMB power spectrum and the lensing potential,\footnote{The lensing power spectrum can also be reconstructed from the CMB temperature and polarization power spectra data alone \cite{Carron:2017vfg} and the result is compatible with $A_L=1$ (see Fig.~3 of \cite{Aghanim:2018eyx}). This further motivates us to look for an extra effect in the primordial power spectrum which mimicks the smoothing effect of lensing.} one can estimate the magnitude of the smoothing effect of lensing $A_L$ \cite{Lewis:2006fu,Calabrese:2008rt}. However, if nothing more than a
power-law inflationary spectrum of adiabatic perturbations and
$\Lambda$CDM are assumed, then the observed power spectrum has
been lensed $10\%$ more than expected \cite{Aghanim:2018eyx}.

\begin{figure}
\begin{tabular}{cc}
\hspace{-4mm}
\includegraphics[width=0.5\columnwidth,valign=m]{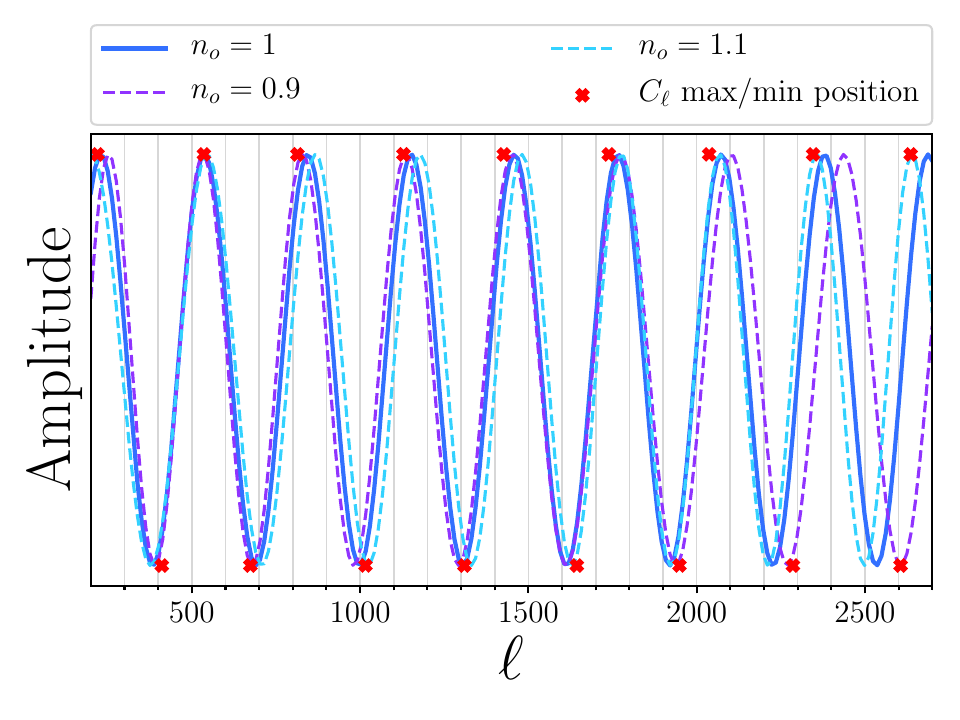}&
    \begin{tabular}{||c | c | c||}
    \hline
    \footnotesize Frequency ($\omega$) & \footnotesize Power-law index ($n_o$) &  \footnotesize Phase ($\varphi$)
    \\ [0.5ex] 
    \hline\hline
     $14.6$  & $1.1$ & $0.5\pi$
     \\[0.5ex] 
    $16.6$ & $1.0$ &  $0.0$
     \\[0.5ex] 
    $18.8$ & $0.9$ & $1.4\pi$\\
    \hline
   \end{tabular}
\end{tabular}
\captionlistentry[table]{parameters}\label{table:parameters}
\captionsetup{labelformat=andtable}
\caption{Left: Oscillatory modulation of the power spectrum with a
constant amplitude. In blue we see the fit for $n_o=1$, in light
blue the one for $n_o=1.1$ and in purple for $n_o=0.9$. The red
crosses are the positions of the maxima and minima of the
acoustic peaks with the amplitude normalized to an arbitrary constant. See
how even though $n_o=1$ offers a fairly good fit, the values of
$n_o=1.1$ and $n_o=0.9$ respectively fit the maxima and the
minima better. The fits to the frequency of the acoustic peaks
have a fixed pivot multipole number at $\ell_*=814$. We note
that the value of $\omega=16.6$ for $n_o=1$ at $k=0.0588 {\rm
Mpc}^{-1}$ agrees with the value $\omega= 14.1$ used by Planck
\cite{Akrami:2018odb} at $k=0.05 \, {\rm
Mpc}^{-1}$. Right: Table containing the fit by eye values for template \eqref{eq:PR} used in Figure \ref{fig:peaksfreq} on the left, the phase already being corrected by $\pi/2$ so that the oscillation is out of phase with the acoustic peaks.  \label{fig:peaksfreq}}
\end{figure}

This tension could conceivably by explained by an oscillatory
modulation of the inflationary power spectrum which is out of
phase with the acoustic peaks. To illustrate this, we introduce the
following fitting to the inflationary power spectrum
\cite{Chen:2011zf,Huang:2016quc,Domenech:2018bnf,Aghanim:2018eyx}:
\begin{align}\label{eq:PR}
\frac{\Delta P_{\cal R}}{P_{{\cal R},0}}= A\left(\frac{k}{k_*}\right)^{n_A}
\sin\left[\omega\left(\frac{k}{k_*}\right)^{n_o}+\varphi\right]
\end{align}
 where the constants $A$, $k_*$, $\omega$, $\varphi$, $n_A$ and $n_o$ respectively are the amplitude, the pivot scale, the frequency, the phase and the power indexes of the $k$ in amplitude and frequency of the oscillation. These constants are ultimately related to parameters of a theoretical model. For example, the fitting form Eq.~\eqref{eq:PR} appears in sudden transitions \cite{Starobinsky:1992ts,Kamionkowski:1999vp,Gong:2001he,Stewart:2001cd,Adams:2001vc,Choe:2004zg,Dvorkin:2009ne,Achucarro:2010da,Adshead:2011jq,Shiu:2011qw,Cespedes:2012hu,Gao:2013ota,Bartolo:2013exa,Achucarro:2014msa}, oscillating heavy fields \cite{Chen:2011zf,Huang:2016quc,Domenech:2018bnf} and trans-planckian modulations \cite{Easther:2005yr} during inflation. Now, we translate it to the multipole number for a rough comparison as
\begin{align}\label{eq:Cl}
\frac{\Delta C_{\ell}}{C_{{\ell},0}}= A\left(\frac{\ell}{\ell_*}\right)^{n_A}
\sin\left[\omega\left(\frac{\ell}{\ell_*}\right)^{n_o}+\varphi\right]\,,
\end{align}
where $C_{{\ell},0}$ is the unlensed power spectrum, we used the relation $\ell\sim k D_A$
($D_A\approx13846\,{\rm Mpc}$ is the comoving angular distance
to the CMB) and as a pivot scale we chose the position of the
third peak $\ell_* \sim 814$ which corresponds to $k_*\sim
0.0588\,{\rm Mpc}^{-1}$. To provide a rough fit to the acoustic
peaks, we first focus on the frequency of the oscillations and
normalize the amplitude to unity. Then we will fit the frequency and phase by eye as we are only interested in the general behaviour. A best fit using CMB data will be studied elsewhere. Since the maxima and minima do
not exactly match a sinusoidal function linear in $\ell$
($n_o=1$), we explore two more possibilities: the power-law
index of the frequency $n_o$ is either $n_o>1$ (fits the maxima)
or $n_o<1$ (fits the minima). See Fig.~\ref{fig:peaksfreq} for a plot of the fits and Table \ref{table:parameters} for the numerical parameters used. This
flexibility in $n_o$ will be important in
Sec.~\ref{sec:features} when we discuss the possible models as
not all models are able to reproduce an exact linear behavior, that is $n_o=1$. We limit ourselves to the case of $n_o=1$ (constant frequency) or very close to it. We leave for future studies different values of $n_o$ in which the oscillation may only fit few consecutive peaks.

\begin{figure}
\includegraphics[width=0.49\columnwidth]{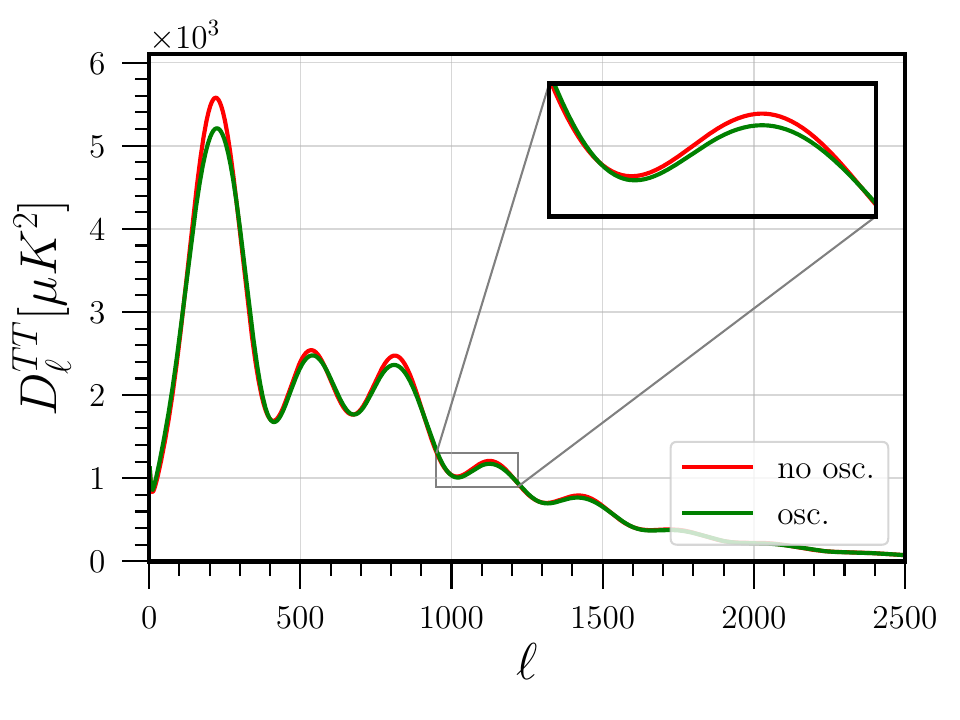}
\includegraphics[width=0.49\columnwidth]{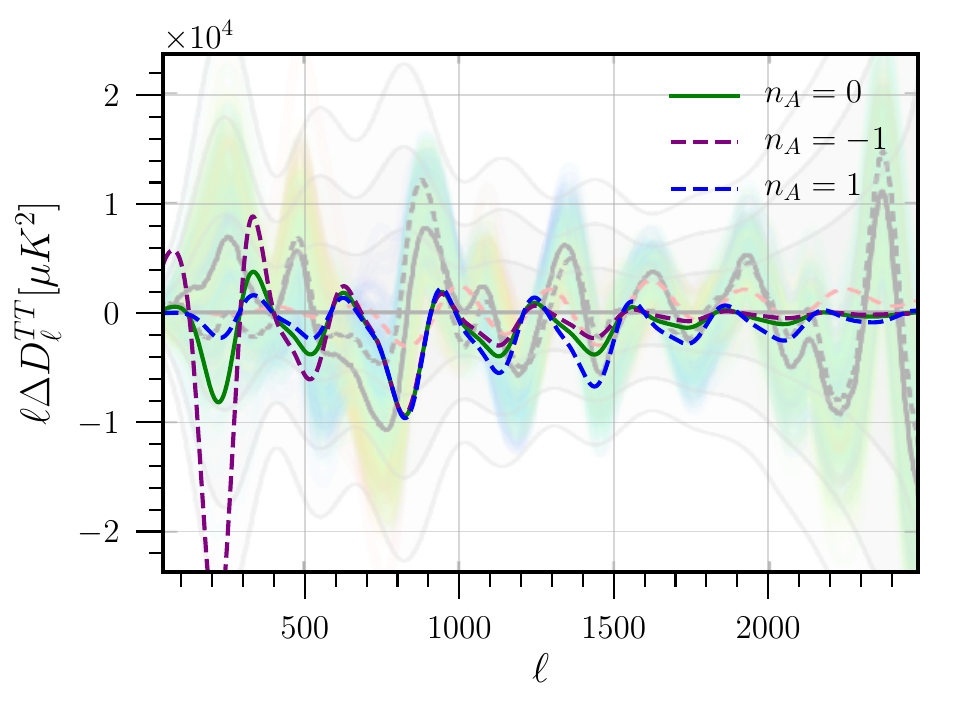}
\caption{CMB lensed power spectrum and residuals for
$\omega=16.6$, $n_o=1$ and $k_*=0.0588{\rm Mpc}^{-1}$
($\ell_*=814$). On the left we compare the lensed power spectrum
with and without the oscillations, respectively in green and
red. We chose $n_A=0$ and $A=0.1$ so that the effects are
clearer. In general, it does not mimick the smoothing effect of
lensing. On the right, we compare the residuals between the
lensed power spectrums (with and without oscillations) for three
values of $n_A=\{-1,0,1\}$ with $A=0.01$ so that all of them
have the same amplitude at $\ell_*$. For easier comparison, we have included in lower opacity Fig. 24 from Ref.~\cite{Aghanim:2018eyx} as a background. In that image, we see in black the residuals of the $\Lambda$CDM model and the gray lines indicate the 1, 2, 3$\sigma$ countours. The green shadows refer to different values of $\Omega_m\,h^2$ which are not relevant for the present discussion. The red-orangish dotted line is the remaining residuals if there were $10\%$ more lensing. Compare our results (blue, green and purple) with the Planck residuals. Note how in the range $\ell\sim 1200-2000$ the two lines (black and red-orangish from Planck) are similar and follow the frequency of the acoustic peaks. Now, we see that the oscillations in Eq.~\eqref{eq:PR} that
could potentially resemble the Planck residuals
\cite{Aghanim:2018eyx} is the one with
$n_A>0$,  since the peaks at large $\ell$ do not decay that fast.   \label{fig:residuals}}
\end{figure}

We computed the effect of the feature \eqref{eq:PR} in the
primordial power spectrum onto the lensed CMB temperature power
spectrum using CLASS \cite{Lesgourgues:2011re,Blas:2011rf}. We
chose $H=100\,h{\rm km/s/Mpc}$, $h = 0.67556$, $T_{\rm
CMB} = 2.7255 {\rm K}$, $\Omega_{\rm b}h^2=0.02$, $\Omega_{\rm
CDM}h^2= 0.12$, $N_{\rm eff}=3.046$, $\Omega_K=0$,
$\Omega_\Lambda=0.69$ and $A_L=1$. For the main power spectrum
we took
 a power-law spectrum $P_{{\cal
R},0}(k)=A_s\left(k/k_{\rm pivot}\right)^{n_s-1}$ with
$A_s=2.2\cdot10^{-9}$, $n_s=0.962$ and $k_{\rm pivot}=0.05 {\rm
Mpc}^{-1}$. For the oscillating feature we use the template \eqref{eq:PR} with the values presented in Table~\ref{table:parameters}. In order to numerically implement a scale dependent amplitude
with CLASS we have introduced an artificial cut-off at $k_c$ in
the power spectrum,  otherwise the spectrum eventually blows up for $n_A\neq0$, so that our power-spectrum reads
\begin{align}
P(k)=P_{{\cal R},0}(k)+\Delta P_{\cal R}(k)\times\left(1 +\tanh\left[\beta\log(k/k_c)\right]\right)/2\,,
\end{align}
and we respectively {used} $k_c=0.001\, {\rm Mpc}^{-1}$ ($\ell\sim 14$)
and $\beta=10$ for $n_A<0$ and $k_c=1{\rm Mpc}^{-1}$ and
$\beta=-10$ for $n_A>0$. For the oscillations in $k$, we further
chose $n_o=1$, $\omega=16.6$ and $A=0.01$, except for the
illustrative case where we 
considered $A=0.1$.  It should be noted that this artificial cut-off introduced in this section will not be necessary when we study concrete inflationary models. The results can be
seen in Fig.~\ref{fig:residuals}. On the left, we plotted an
illustrative case with $A=0.1$ and we see that the oscillations
do not exactly mimick the effect of smoothing (compare with
Fig.~\ref{fig:lensed}) but it could be enough at the level of
the residuals. On the right, we plotted the residuals between
the lensed temperature power specturm with and without the
oscillation for three different cases: $n_A=\{-1,0,1\}$. As one
can see the residuals for $n_A=1$ are the ones that best resemble
the residuals from the Planck 2018 analysis
\cite{Aghanim:2018eyx}, in particular for $\ell>814$. From the
data, it seems that $n_A>0$ will be preferred but it is unclear how
one could generate a growing feature, rather a decaying feature
is expected from inflationary dynamics as we shall see in the
next section. It should be noted that the apparent power loss in Fig.~\ref{fig:residuals} is due to the enhancement of the contribution of density peaks (corresponding to peaks in the power spectrum) relative to velocity (Doppler) peaks (corresponding to troughs in the power spectrum) to temperature fluctuations \cite{Hu:1995em}. Thus, for the same initial conditions, the peaks in the power spectrum have a larger contribution from the oscillatory feature than the troughs.

\section{Features during inflation}\label{sec:features}

In this section, we review the computation of the primordial
power spectrum when there is a feature, e.g., sharp transition or
oscillations, in the background evolution. In fact, such features are rather common in extensions of the simplest models of inflation. For example, it could be due to wiggles in the inflaton potential, inspired from axion monodromy in string theory \cite{Silverstein:2008sg,Flauger:2009ab}, or sudden turns in the trajectory in field space in multi-field inflation \cite{Gao:2013ota}, which would excite the heavy modes. These features will affect the background dynamics during inflation and the effects will be imprinted in modifications of the slow-roll parameters and/or the sound speed of perturbations.  Now, for simplicity, we
take an effective single field approach
\cite{Achucarro:2010jv,Shiu:2011qw,Achucarro:2012sm,Bartolo:2013exa,Palma:2014hra}
and we study the resulting oscillatory modulation of the power
spectrum. Our starting point is then the Mukhanov-Sasaki equation
for the canonically normalized curvature perturbation
\cite{Kodama:1985bj,Mukhanov:1990me}:
\begin{align}\label{eq:MS}
u_k''+\left(c_s^2k^2-\frac{z''}{z}\right)u_k=0\, ,
\end{align}
where $z^2\equiv2a^2\epsilon M_{\rm pl}^2/c_s^2$, $a$ is the scale factor, $\epsilon\equiv-\dot H/H^2$, $H\equiv\dot a/a$, $c_s^2$ is the sound speed of propagation, $\dot\,\,\equiv d/dt $ where $t$ is the cosmic time,  $'\,\,\equiv d/d\tau $ where $d\tau=dt/a$ is the conformal time and $u\equiv z{\cal R}$ with ${\cal R}$ being the comoving curvature perturbation. We then consider the effect of a deviation in a de-Sitter inflationary background by introducing \cite{Achucarro:2014msa} 
\begin{align}
v_k\equiv\sqrt{2kc_s}u_k\quad{\rm and}\quad f\equiv2\pi z c_s^{1/2}\xi
\end{align}
where $d\xi=c_sd\tau$. With these redefinitions, Eq.~\eqref{eq:MS} becomes \cite{Achucarro:2014msa} 
\begin{align}\label{eq:MS2}
\frac{dv_{k}}{d\xi}+\left(k^2-\frac{2}{\xi^2}\right)v_{k}=\frac{1}{\xi^2f}\left(\frac{d^2f}{d\ln\xi^2}-3\frac{df}{d\ln\xi}\right)v_k\,.
\end{align}
Treating the right hand side as a perturbation one can solve the differential equation by the Green's function method at leading order in $f$ by \cite{Gong:2001he,Joy:2005ep,Gong:2005jr,Bartolo:2013exa,Achucarro:2014msa}
\begin{align}\label{eq:deltaP}
\frac{\Delta P_{\cal R}(k)}{P_{{\cal R},0}}=\int_{-\infty}^\infty d\ln\xi \,W(k\xi)\,\left(\frac{2}{3}\frac{d^2\ln f}{d\ln\xi^2}-2\frac{d\ln f}{d\ln\xi}\right)
\end{align}
where we already used the dS approximation, i.e. $v_{k}=\left(1-\frac{i}{k\xi}\right){\rm e}^{-ik\xi}$, we defined 
\begin{align}
P_{\cal R}(k)\equiv\frac{k^3}{2\pi^2}|v_k|^2\quad{\rm where}\quad P_{{\cal R},0}(k)\equiv\frac{1}{8\pi^2}\frac{H^2}{\epsilon c_sM_{\rm pl}^2}\,,
\end{align}
and
\begin{align}
W(k\xi)\equiv\frac{3}{2}\left[\frac{\sin 2k\xi}{(k\xi)^3}-\frac{2\cos2k\xi}{(k\xi)^2}-\frac{\sin 2k\xi}{k\xi}\right]\,.
\end{align}
This will be our starting point in the following discussions. Any feature during inflation will be contained in the function $f(\xi)$. Thus, once we know the type of feature we can compute its effect in the power spectrum by using Eq.~\eqref{eq:deltaP} through modifications of the slow-roll parameters or the propagation speed. Interestingly, one can invert this relation and find the feature given a power-spectrum modulation as in Ref.~\cite{Joy:2005ep} (also see Ref.~\cite{Durakovic:2019kqq} for a more recent approach). At this point, we could find the change in the background that would lead to the desired feature in the power spectrum. However, we will be more interested in the physical model behind. We will analytically compute three different regimes: $(i)$ fast oscillating features, $(ii)$ slow oscillating features and $(iii)$ sharp features. Here we do not seek to join any of these three regimes, rather we are interested to see if the desired oscillations in the power spectrum fall in any of these three categories.

\subsection{Fast Oscillating feature\label{subsec:fast}}

We begin to review the effects of an oscillating modulation of the background where its frequency is higher than the expansion rate. 
This could be either induced by an oscillatory modulation of the inflaton's potential or by the oscillations of an extra massive field. For the moment, we will assume that the oscillations vary in amplitude and frequency and that $c_s=1$. Thus, in practice we have that
\begin{align}\label{eq:deltaPfast}
\frac{\Delta P_{\cal R}(k)}{P_{{\cal R},0}}=\frac{2}{3}\int_{-\infty}^\infty d\ln\tau \,W(k\tau)\,\frac{\Delta\left(\tau\right)}{H^2}
\end{align}
where we assumed that the frequency of the oscillation, say $\Omega$, in the function $f$ is $\dot\Omega/H=\delta_\Omega\Omega\gg1$ then only the highest time derivative dominates and so
\begin{align}
\frac{\Delta(\tau)}{H^2}\equiv\frac{d^2\ln f}{d\ln\tau^2}=-C(\tau)\cos\left[\Omega(\tau)+\varphi\right]\,,
\end{align}
with $\varphi$ being an arbitrary phase. For simplicity, we will further assume that
\begin{align}
C(\tau)=C_r\left(\frac{a}{a_{r}}\right)^{\delta_C}\qquad{\rm and}\qquad \Omega(\tau)=\Omega_r\left(\frac{a}{a_{r}}\right)^{\delta_\Omega}\,,
\end{align}
where $a_r$ is the scale factor at onset of the resonance, $C_r$, $\Omega_r$, $\delta_C$, $\delta_\Omega$ are constants and we require $\dot\Omega/H=\delta_\Omega\Omega\gg1$. Then, we can use the saddle point approximation for subhorizon scales ($k\tau\gg 1$) in Eq.~\eqref{eq:deltaPfast} at $k=\delta_\Omega aH\Omega$ to find that the correction to the power-spectrum is given by
\begin{align}\label{eq:fastoscillating}
\frac{\Delta P_{\cal R}(k)}{P_{{\cal R},0}}\approx\sqrt{2\pi}\frac{C_r\left(\left|\delta_\Omega\right|\Omega_r\right)^{-3/2}}{\sqrt{|1-\delta_\Omega|}}\left(\frac{k}{k_r}\right)^\frac{\delta_C-3\delta_\Omega/2}{1+\delta_\Omega}\sin\left[\Omega_r\left(1+\delta_\Omega\right)\left(\frac{k}{k_r}\right)^\frac{\delta_\Omega}{1+\delta_\Omega}+\tilde\varphi\right]
\end{align}
where $k_r=\delta_\Omega\Omega_r a_r H_r$ and $\tilde
\varphi=\varphi\mp\frac{3\pi}{4}$ where $-$ is for
$\delta_\Omega>1$ and $+$ for $\delta_\Omega<1$.

\begin{figure}
\includegraphics[width=0.49\columnwidth]{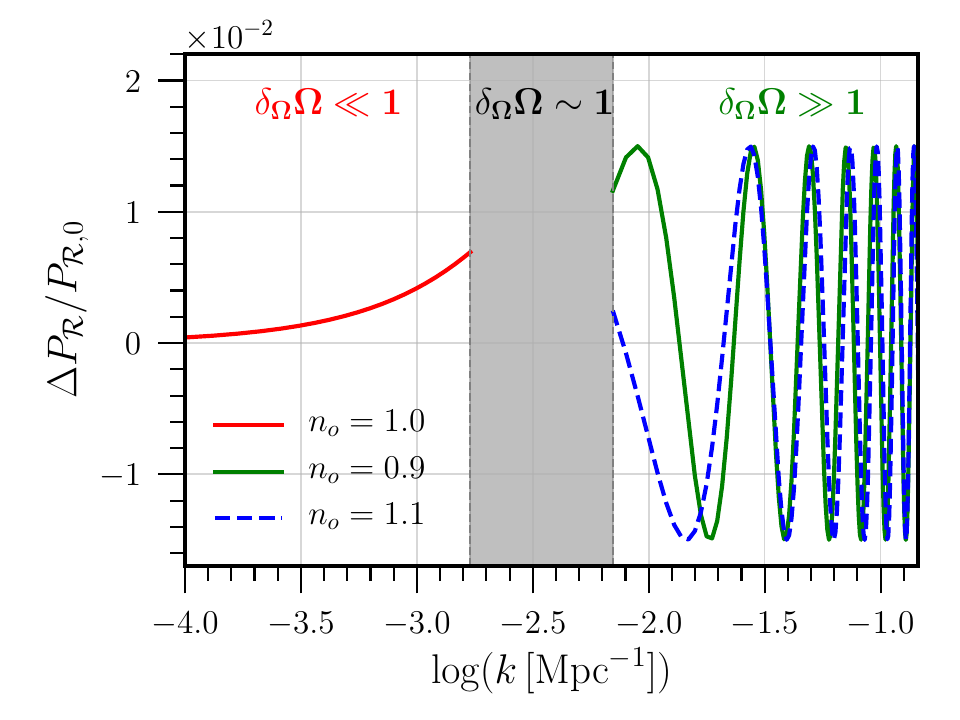}
\includegraphics[width=0.49\columnwidth]{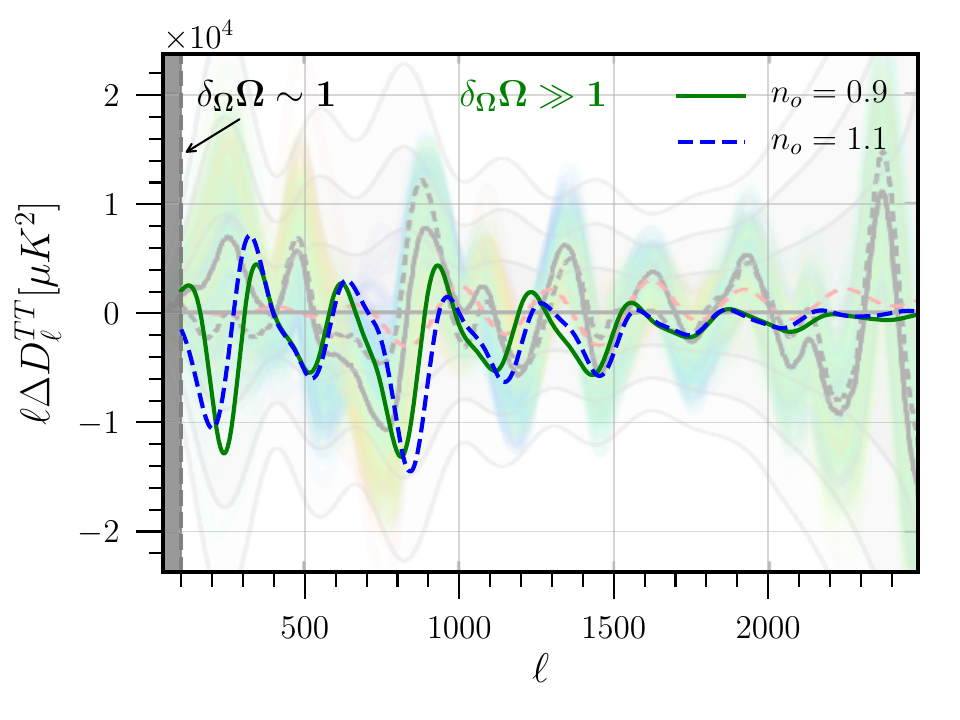}
\caption{Left: Power spectrum modulation for the fast ($\delta_\Omega \Omega\gg1$) and slow ($\delta_\Omega \Omega\ll1$) oscillating regimes separated by an intermediate regime where numerical computations are needed (from $\delta_\Omega \Omega\sim1/2$ to $\sim 2$). For the fast oscillating regime we chose $\delta_\Omega=9$ (green line), $\delta_\Omega=-11$ (dashed blue line) both cases with $\delta_C=3\delta_\Omega/2$ in Eq.~\eqref{eq:fastoscillating} which corresponds to $n_A=0$ and $n_o=0.9$ and $n_o=1.1$ respectively in Eq.~\eqref{eq:PR}. For the slow oscillating regime (red line) we chose $\delta_\Omega=1$ and $\delta_B=0$ in Eq.~\eqref{eq:slow} which corresponds to $n_A=0$ and $n_o=1$ in Eq.~\eqref{eq:PR}. In all cases we used Tab.~\ref{table:parameters} for the values of the frequency and phase. Right: Residuals of the lensed power spectrum for the same cases than in the left figure. For comparison, we included Fig.~(24) of Planck analysis \cite{Aghanim:2018eyx}. In black we have the residuals of the $\Lambda$CDM model, gray lines are the 1, 2, 3$\sigma$ countours and the green shadows refer to different values of $\Omega_m\,h^2$. The red-orangish dotted line is the remaining residuals if there were $10\%$ more lensing.\label{fig:example2}}
\end{figure}

In order to compare with the data, we rewrite the parameters in Eq.~\eqref{eq:fastoscillating} in terms of the template Eq.~\eqref{eq:PR}. Comparing them at a scale $k=k_*$ we find that for the frequency we have
\begin{align}
\Omega_r={\omega}{\left|1-n_o\right|}\left(\frac{k_r}{k_*}\right)^{n_o}
\quad{\rm ,}\quad
\delta_\Omega=\frac{n_o}{1-n_o},
\end{align}
and for the amplitude
\begin{align}
C_r=\frac{A}{\sqrt{2\pi}}\left(\omega n_o\right)^{3/2}\left(\frac{k_r}{k_*}\right)^{3n_o/2}\sqrt{\left|\frac{1-2n_o}{1-n_o}\right|}\quad{\rm ,}\quad
\delta_C=\frac{n_A+3n_o/2}{1-n_o}\,.
\end{align}
First of all, we see that an exact linear dependence in $k$ for
the frequency, as needed to fit all the acoustic peaks, is not possible: as $n_o\to 1$ we have that
$\delta_\Omega\to\pm \infty$. This is because one would need
that the resonance with every mode function, which has a
frequency of $2k\xi$, occurred at the same time for all the
modes, i.e. at $\xi=\xi_r$. Nevertheless, the best one can do
without considering a sharp feature is that the oscillating
source term oscillates so fast that the resonances occur almost
at the same time. For this reason, we may consider that
$n_o=\{0.9,1.1\}$ and the fit is still reasonable, see Fig.~\ref{fig:example2}. However, while they might be give feasible residuals to mimick the smoothing effect of lensing, they are unsatisfactory from a theoretical point of view.

First of all, it must be seen whether the assumption $\delta_\Omega\Omega\gg1$
holds for the range of interest. Plugging in some numbers (see
Tab.~\ref{fig:peaksfreq}), we find that
\begin{align}\label{eq:freq}
\delta_\Omega\Omega_r={n_o\omega}\left(\frac{k_r}{k_*}\right)^{n_o}\approx 1.7\times 10^{-1}\frac{n_o\omega}{16.6}\left(\frac{k_r}{10^{-2}k_*}\right)^{n_o}\,,
\end{align}
 where we used the fact that $n_o\approx 1$. We see that, using the values of the frequencies in Tab.~\ref{table:parameters}, the condition $\delta_\Omega\Omega\gg 1$ breaks down at around $k\sim0.0035$ ($\ell\sim49$) and so this kind of resonance could explain the anomaly down to small $\ell$. However, to be able to predict what occurs for $\ell<49$, as it enters the observational window, we need a full detailed model. For instance, it would difficult to imagine a model where the modulation suddenly started at $k\sim0.0035$ since it would be accompanied by a sharp feature which usually has an amplitude higher relative to the fast oscillating regime and quite model dependent \cite{Chen:2014cwa}.

Second and most important, we need to see if it is theoretically viable. To be close to a constant frequency we chose $n_o=\{0.9,1.1\}$ in Eq.~\eqref{eq:PR} which correspond to $\delta_\Omega=\{9,-11\}$ in Eq.~\eqref{eq:fastoscillating}. This implies that the value of the frequency respectively increases or decreases 4 orders of magnitude in $1$ e-fold. In this respect, it is difficult to conceive what kind of model could sustain such growth or decay for more than the $3$ e-folds required. For example, in models where the feature is generated by an extra oscillating massive field like in Refs.~\cite{Chen:2011zf,Huang:2016quc,Domenech:2018bnf}, the frequency of the oscillation is associated to the mass of the field. In this case, either the energy density of the field increases too fast with the mass ($\delta_\Omega=9$) or it decreases so fast ($\delta_\Omega=-11$) that initially it should have had an astoundingly big mass and energy density. See App.~\ref{app:models} for a detailed comparison with such models. It should be noted that perhaps it may be possible to build such a model but we consider the amount of fine tuning required to be unreasonable. For these reasons, we will explore other simpler cases.

We thus disregard the fast oscillating case as a theoretically viable explanation for the $A_L$ anomaly. Nevertheless, it must be seen whether relaxing the assumption that the frequency be almost constant helps finding a viable model which resembles the lensing effect in a narrower multipole number range.


\subsection{Slow Oscillating feature\label{subsec:slow}}

In this subsection, we study the opposite case where the oscillating modulation is slowly varying. Thus, this feature will simply act as a modulation of the background and we can estimate its effects onto the power spectrum using the $\delta N$ formalism\footnote{Since we are interested in the super-horizon limit, it could also be estimated from Eq.~\eqref{eq:MS} or \eqref{eq:MS2} with the approximation that $z$ is almost constant and then matching the solution at horizon crossing \cite{Hu:2011vr}. This means that \eqref{eq:deltaP} is unnecessary for the slow oscillating feature; $\Delta P_{\cal R}$ is defined as difference between $P_{{\cal R},0}$ with and without oscillation.} \cite{Starobinsky:1986fxa,Salopek:1990jq,Sasaki:1995aw}.  We have
\begin{align}
\delta N=\frac{\partial N}{\partial \phi}\delta\phi+...
\end{align}
where $\frac{\partial N}{\partial \phi}=-\frac{H}{\dot\phi}$ is to be evaluated at horizon crossing. In general one has that $\langle\delta\phi\delta\phi\rangle=H/(2\pi c_s)$ and so the power spectrum is given by
\begin{align}
 P_{{\cal R}}=\left(\frac{\partial N}{\partial \phi}\right)^2\langle\delta\phi\delta\phi\rangle=\frac{1}{8\pi^2}\frac{H^2}{\epsilon c_sM_{\rm pl}^2}\,.
\end{align}
The effect of any slow varying modulation of the background can be computed in this way, evaluated at horizon crossing $k=aH$. Now, let us assume that the modulation of the background results in a modulation of the slow-roll parameter given by
\begin{align}
\epsilon=\epsilon_0\left(1-B(\tau)\sin\left[\Omega(\tau)+\varphi\right]\right)\,.
\end{align}
Assuming for simplicity that
\begin{align}
B(\tau)=B_r\left(\frac{a}{a_{ r}}\right)^{\delta_B}\qquad{\rm and}\qquad \Omega(\tau)=\Omega_r\left(\frac{a}{a_{ r}}\right)^{\delta_\Omega}\,,
\end{align}
where $B_r$, $\Omega_r$, $\delta_B$ and $\delta_\Omega$ are constants, together with the requirement that $\delta_\Omega\Omega\ll1$, we find that the modulation of the power-spectrum reads
\begin{align}\label{eq:slow}
\frac{\Delta P_{\cal R}(k)}{P_{{\cal R},0}}\approx B_r\left(\frac{k}{k_{ r}}\right)^{\delta_B}\sin\left[\Omega_r\left(\frac{k}{k_{ r}}\right)^{\delta_\Omega}+\varphi\right]\,.
\end{align}
Comparing this result with the template \eqref{eq:PR} we find
\begin{align}
\Omega_r=\omega \frac{k_r}{k_*}\quad,\quad \delta_\Omega =n_o\quad,\quad B_r=A \quad{\rm and}\quad \delta_B=n_A\,.
\end{align}
This time, we have that for $\delta_\Omega=n_o=1$ at
$k_r=10^{-2}k_*$, the oscillations are slowly varying,
i.e. $\delta_\Omega\Omega_r\approx 1.7\times 10^{-1}$. However,
since the frequency is increasing with time as a power-law of
$a$, the approximation of $\delta_\Omega\Omega\ll1$ will break down at around $\ell\sim49$. This is clearly not enough to explain the lensing anomaly at $\ell\sim 1000-2000$.
 We will not attempt to join the slow and fast regimes of Sec.~\ref{subsec:fast}, as it is not clear how to go from $\delta_\Omega\Omega\ll1$ to $\delta_\Omega\Omega\gg1$ smoothly and, furthermore, numerical computations would be needed.

\subsection{Sharp feature\label{subsec:sharp}}

When one considers a sharp feature, the exact shape of the
modulation is very model dependent
\cite{Starobinsky:1992ts,Adams:2001vc,Choe:2004zg,Gong:2005jr,Dvorkin:2009ne,Achucarro:2010da,Adshead:2011jq,Shiu:2011qw,Gao:2013ota,Bartolo:2013exa,Palma:2014hra}. Nevertheless,
let us consider the simplest example where the sharp feature is
a discontinuity, e.g. a step in the slope of the potential,
which will result in a Dirac delta $\delta(\xi-\xi_f)$ in
Eq.~\eqref{eq:deltaP}, as the slow roll parameter $\epsilon$ in $f(\xi)$ is proportional to the first derivative of the potential. In that case, the frequency of the resulting
oscillation will be proportional to $2k\xi_f$
\cite{Gong:2001he,Joy:2005ep,Gong:2005jr}. A quick exercise
tells us that if we require $\omega=16.6$ the transition
happened at $k_*\xi_r\approx 8.3$, that is the transition
happened $2$ e-folds before our pivot scale at around
$k_f\approx0.007 {\rm Mpc}^{-1}$ or $\ell_r\approx
98$. Furthermore, the phase of the oscillation depends on
whether the step around  {$\xi=\xi_f$} is odd (e.g. a hyperbolic tangent)
\cite{Starobinsky:1992ts,Choe:2004zg,Gong:2005jr,Bartolo:2013exa}
or even (e.g. a gaussian bump) \cite{Achucarro:2014msa}. To
understand that it is useful to integrate by parts
Eq.~\eqref{eq:deltaP} arriving at
\begin{align}\label{eq:deltaP2}
\frac{\Delta P_{\cal R}(k)}{P_{{\cal R},0}}=-2\int_{-\infty}^\infty d\ln\xi \,\left(W(k\xi)+\frac{1}{3}\frac{dW(k\xi)}{d\ln\xi}\right)\,\frac{d\ln f}{d\ln\xi}\,.
\end{align}
If the step is sharp enough only the neighborhood of the transition will contribute to the integral. If the step is odd around $\xi=\xi_f$, its derivative is even and so the even function $\cos(2k\xi_f)$ survives asymptotically in $k$. Instead, if the step is even around  $\xi=\xi_f$, then its derivative is odd and the odd function $\sin(2k\xi_f)$ remains.

\begin{figure}
\includegraphics[width=0.49\columnwidth]{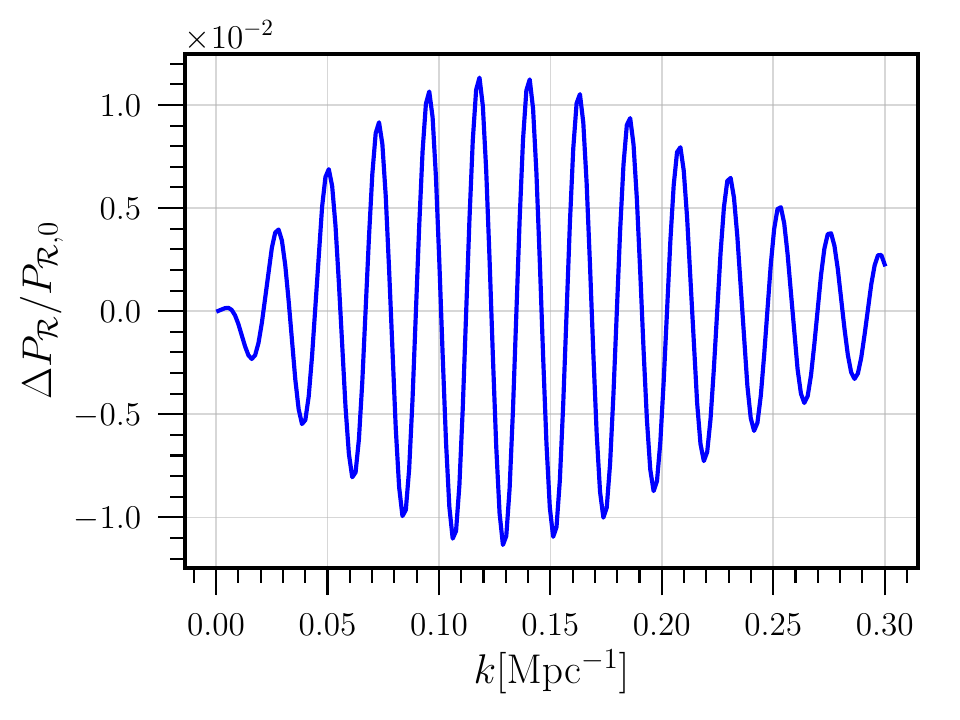}
\includegraphics[width=0.49\columnwidth]{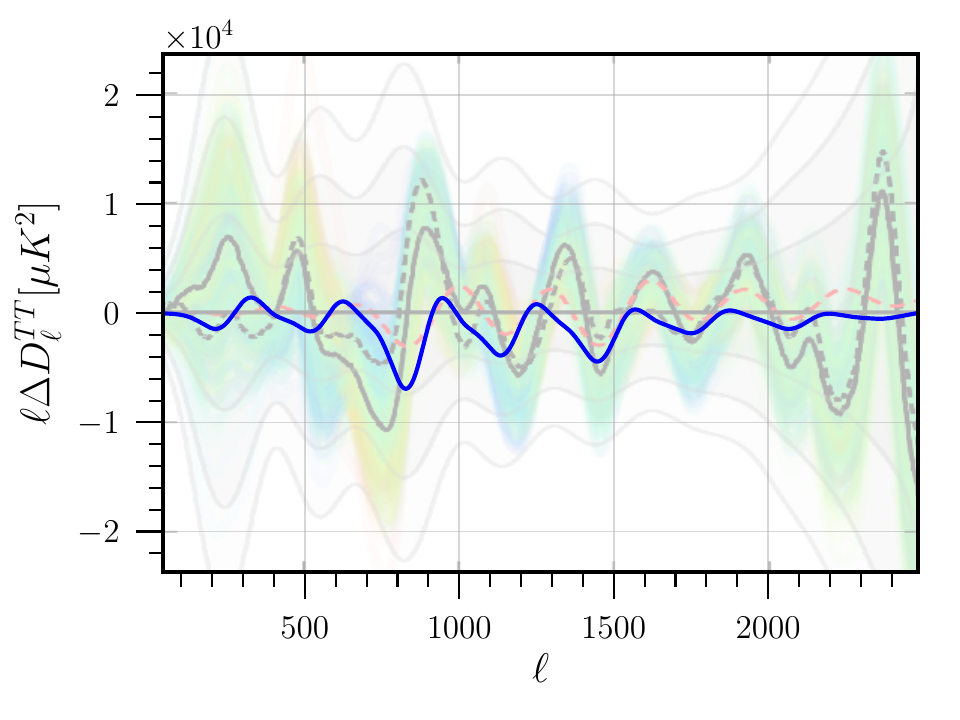}
\caption{Power spectrum modulation \eqref{eq:final} (left) and residuals of the lensed power spectrum (right) for a bump \eqref{eq:bump} in $c_s^2$ with height $B=-0.015$ and sharpness $\beta_s=25$. Note that on the right the residuals are similar to those in the Planck analysis \cite{Aghanim:2018eyx} Fig.~(24), which we included for easier comparison. Again, we have in black the residuals of the $\Lambda$CDM model, the gray lines are the 1, 2, 3$\sigma$ countours and the green shadows refer to different values of $\Omega_m\,h^2$. The red-orangish dotted line is the remaining residuals if there were $10\%$ more lensing. Compare our result (blue) with the Planck residuals (black and red-orangish). Note also that although the frequency of our resulting oscillations have a similar pattern, the amplitude of the Planck residuals are shifted upwards compared to our result.\label{fig:example}}
\end{figure}

We present now a concrete example. The simplest case is a bump in the sound speed at $\tau=\tau_f$ with height $B$ and sharpness $\beta_s$ given by \cite{Achucarro:2014msa}
\begin{align}\label{eq:bump}
c_s^2=1+B{\rm e}^{-\beta_s^2\log^2\left[\tau/\tau_f\right]}\,.
\end{align}
Note that if $B>0$ the sound speed of scalar perturbations is slightly superluminal.\footnote{In Ref.~\cite{Achucarro:2014msa} it is assumed that $B<0$ throughout the paper but for our purposes we shall consider $B>0$ as well.} Although it does not pose any causality problems \cite{Babichev:2007dw}, it may obstruct the UV completion of a quantum Lorentz-invariant theory \cite{Adams:2006sv}. Nevertheless, the superluminality could be compensated by introducing a $c^2_{s,0}$ as a common factor in Eq.~\eqref{eq:bump} with $(1+B)c^2_{s,0}\leq1$.
Now, integrating Eq.~\eqref{eq:deltaP2} under the assumption that the step is sharp {($\beta_s\gg1$)} yields \cite{Achucarro:2014msa} (see also App.~\ref{App:spectrum})
\begin{align}\label{eq:final}
\frac{\Delta P_{\cal R}(k)}{P_{{\cal R},0}}=B\sqrt{\pi}\frac{k\tau_f}{\beta_s}{\rm e}^{-\frac{k^2\tau_f^2}{\beta_s^2}}\left\{\sin(2k\tau_f)+\frac{\cos(2k\tau_f)}{k\tau_f}-\frac{1}{2}\frac{\sin(2k\tau_f)}{(k\tau_f)^2}\right\}\,,
\end{align}
where we neglected terms $O(\beta_s^{-2})$ and note that $\Delta
P_{\cal R}\to0$ when $k\to0$ so that there are no spurious
super-horizon modes. To estimate the magnitude of the feature, note that the gaussian modulation has a maximum at
$k\tau_f=\beta_s/\sqrt{2}$ and so to have an amplitude of $10\%$
one needs $|B|\approx0.013$.  It should be noted that the tensor-to-scalar ratio , which is proportional to $c_s$ \cite{Mukhanov:2005bu}, is barely affected as the sound speed only changes by $10\%$. Furthermore, the adiabatic condition $s\equiv \dot c_s/Hc_s$ is always satisfied and its maximum value is $|s|=|B|\beta_s/\sqrt{2e}\approx 0.16$ (for $|B|=0.015$ and $\beta_s=25$). We have plotted the oscillatory
feature in the primordial power spectrum and the residuals of
the lensed power spectrum in Fig.~\ref{fig:example}.  {Note how the frequency of the resulting oscillatory pattern follows that of Planck \cite{Aghanim:2018eyx} in the range $\ell\sim 1200-2000$, although the residuals from Planck are slightly shifted upwards.}

Regarding the non-gaussianities we can borrow the results from
Ref.~\cite{Achucarro:2014msa} (see also App.~\ref{App:trispectrum}) at the equilateral limit and find that the equilateral non-gaussianity peaks at
$K\tau_f=\sqrt{6}\beta_s$ with amplitude
\begin{align}
f_{NL}^{\rm eq}\approx -0.27\,B\beta_s^2\,.
\end{align}
A similar calculation for the trispectrum evaluated at its peak ($K\tau_f=\sqrt{10}\beta_s$) in the equilateral configuration (see App.~\ref{App:trispectrum}) yields
\begin{align}
g^{\rm eq}_{NL}\approx -0.1\,B\beta_s^4\,.
\end{align}
The Planck results on non-gaussianities \cite{Ade:2015ava}
yield that $f_{NL}^{\rm eq}=-4\pm 43$. Thus, if $|B|\sim 10^{-2}$ we
see that we need $\beta_s<60$ to fall within the
bounds. Regarding local (squeezed shape) non-gaussianity its magnitude is at least suppressed
by $1/\beta_s$ with respect to $f_{NL}^{\rm eq}$ \cite{Achucarro:2012fd,Mooij:2016dsi} and therefore
we easily fall within Planck constraints, i.e. $f_{NL}^{\rm
loc}=0.8\pm5.0$.\footnote{For instance, using the results from \cite{Achucarro:2012fd} we have that $f^{\rm loc}_{NL}=-\frac{5}{12}\frac{d}{d\ln k}\frac{\Delta P_{\cal R}}{P_{{\cal R},0}}\approx -\frac{5}{6}\beta_s \frac{\Delta P_{\cal R}}{P_{{\cal R},0}}$ and plugging in the same numbers, i.e. $|B|\sim 10^{-2}$ and $\beta_s<60$ we have that $|f^{\rm loc}_{NL}|\lesssim0.5$.}  Note that these constraints are looser if one
allows for a scale dependence in the non-gaussianity
\cite{Ade:2015ava}. Furthermore, the constraints on the trispectrum are roughly \cite{Ade:2015ava} $g_{NL}<10^5-10^6$ and for $|B|\sim 10^{-2}$ it is sufficient that $\beta_s<100$. Thus, as we can see in
Fig.~\ref{fig:example} a bump in $c_s^2$ with $B=-0.015$ and
$\beta_s=25$ reproduces quite well the residuals in the Planck
analysis \cite{Aghanim:2018eyx} and is well within the bispectrum and trispectrum bounds. The development of a specific
model that can produce such a bump in $c_s^2$ is left for future
work, although it seems possible to build such a model using a spectator scalar field \cite{Nakashima:2010sa}. Before ending this section, it is worth saying that no trans-planckian modulation \cite{Easther:2005yr} could mimick the smoothing effect of lensing. This is due to the fact that the value of the frequency for trans-planckian modulations depends on the initial conditions but the frequency required to explain the anomaly correspond to a scale in the last few e-folds of inflation.


\section{Conclusions}\label{sec:conclusions}

The latest analysis of the cosmic microwave background by the
Planck team \cite{Aghanim:2018eyx} suggests that at the $2\sigma$
confidence level there is $10\%$ more lensing than predicted by
$\Lambda$CDM. If not a statistical fluke, one suggested
explanation for the extra lensing is that there is new physics
that mimicks the smoothing effect of lensing
\cite{Akrami:2018odb}. Here we studied what could have generated
these oscillations in the power spectrum during inflation. We
first considered an effective single field approach, where the
effects of a sharp transition or an oscillatory modulation in
the background can be studied phenomenologically
\cite{Shiu:2011qw,Achucarro:2012sm,Bartolo:2013exa,Palma:2014hra}. In
this way, we divided the analysis between rapid/slow (compared to
the expansion rate) oscillatory modulations and sharp
transitions. 

We have found that for rapid oscillatory modulations
\cite{Chen:2011zf,Huang:2016quc,Domenech:2018bnf}, it is not
possible to obtain an exact linear $k$ dependence in the
frequency of the power-spectrum's oscillations since the
modulation should oscillate infinitely fast or be a sharp
feature. Nevertheless, an almost linear dependence can be
obtained for very fast oscillatory modulations. Unfortunately,
when compared with the data one needs a frequency which is
slowly varying for large scales ($\ell<50$) and rapidly varying for
$\ell>50$. We also showed that if the oscillations were caused by
an oscillating heavy field, then the mass of the field would
have been smaller than Hubble at some point in the range of
interest. Thus, this sort of feature cannot explain an
oscillation over the whole range of $\ell$ covered by the Planck
data. We discussed that the possibility of starting the
oscillation at $\ell>50$ is not feasible since it would be
accompanied by a sharp feature which is normally larger than the
oscillatory feature. 

On the other hand, we have analyzed the case of slowly oscillating modulations of the background and we have found that it is possible to find a model where the frequency of the oscillatory feature is linear in $k$. In this case, there is no resonance occurring and so the frequency must evolve inversely proportional to the conformal time so that at horizon crossing ($-k\tau\approx1$) yields a linear dependence in $k$. However, when compared with the data and in agreement with the results of fast oscillatory modulations, this feature could only explain a linear oscillation for $\ell<50$ which is not of interest for our work.

Motivated by our results, we have studied sharp transitions
within
an effective single field theory for sharp features
\cite{Shiu:2011qw,Achucarro:2012sm,Bartolo:2013exa,Palma:2014hra}. When
the feature is sharp all the modes are excited at the same time
(say $\tau=\tau_f$) and so the resulting oscillatory feature has
a frequency of $2k\tau_f$. If that is the case, we needed that
the sharp feature occurred at scales inside the observational
window, around $\ell\sim 98$. Although sharp features are
very model dependent
\cite{Starobinsky:1992ts,Adams:2001vc,Choe:2004zg,Gong:2005jr,Dvorkin:2009ne,Achucarro:2010da,Adshead:2011jq,Shiu:2011qw,Gao:2013ota,Bartolo:2013exa,Palma:2014hra},
we see that in general terms when the sharp feature is even
\cite{Achucarro:2014msa}, e.g. a bump,
the oscillations with the right frequency
are in phase with the acoustic
peaks. We have presented an example capable of reproducing the
desired oscillatory modulation of the primordial power spectrum
times a damping function \eqref{eq:final}. This example consists
of a bump in the sound speed given by Eq.~\eqref{eq:bump}.  Moreover, we have shown that this model can satisfy the bounds to the bispectrum and trispectrum.

We thus conclude, on one hand, that the $A_L$ anomaly in the CMB
temperature power spectrum  could potentially be explained by a
bump in the sound speed of scalar perturbations,  although a detailed comparison with the data would be needed. We presented
the residuals in Fig.~\ref{fig:example} and they are similar in frequency to
the results presented in
Ref.~\cite{Aghanim:2018eyx}.  In the future, measurements of
baryon acoustic oscillations might be employed in combination
with the CMB \cite{zeng:2018ufm} to test this
explanation. In the standard scenario, the Fourier wavenumbers
for the peaks in the late-time matter power spectrum are shifted
relative to those for peaks in the radiation density at CMB
decoupling \cite{eisenstein:1997ik}, a result of the fact that
the late-time growing mode maps at early times to a combination
of the growing and decaying modes.  The relative phases of the
acoustic and primordial oscillations will thus be different in
the {baryon acoustic oscillations} (BAO) than they are in the CMB. It will be interesting do this analysis with high precision polarization data such as CMB-S4. Furthermore, another probe of this model would be to look for correlated features in the primordial spectra \cite{Achucarro:2012fd,Gong:2014spa,Chen:2016vvw,Finelli:2016cyd,Ballardini:2016hpi,Ballardini:2017qwq,Palma:2017wxu,Gong:2017vve}. On the other hand, we have
shown also that it is difficult that oscillating features in the
power spectrum which are linear in $k$ (or almost linear) are
generated during inflation from an oscillatory modulation of the
background and that  could explain the $A_L$ anomaly at the same
time. However, it has to be seen if there is any possibility for general multi-field inflationary trajectories as in Ref.~\cite{Gao:2013ota}.

\section*{Acknowledgments}

We would like to thank J-O.~Gong, E.~Kovetz, J.~Mu{\~n}oz, R.~Saito, P.~Shi, J.~Takeda and T.~Tenkanen for very useful discussions and comments on the draft. This work was partially supported by DFG Collaborative Research
center SFB 1225 (ISOQUANT)(G.D.). G.D. acknowledges support from
the Balzan Center for Cosmological Studies Program during his
stay at the Johns Hopkins University. G.D. also thanks the Johns
Hopkins University cosmology and gravity groups for their
hospitality.

\appendix

\section{Model building\label{app:models}}

In this section we give the phenomenological parameters in Sec.~\ref{sec:features} in terms of particular models. We will first consider a two-field model in which the heavy field is excited and oscillates around the minimum of its potential. This could be a particular realization of the case studied in Sec.~\ref{subsec:fast} as the heavier the field the faster the oscillations. In the second example, we will consider that the inflaton's potential has an oscillatory modulation superimposed. This could be an example for either fast or slow oscillations depending on the model parameters and could be used in Secs.~\ref{subsec:fast} and \ref{subsec:slow}.

\subsection{Non-standard clock signal}

Here we review the model studied in \cite{Domenech:2018bnf} which is a generalized version of the standard clock model proposed in Ref.~\cite{Chen:2011zf}. The action is given by
\begin{align}\label{eq:action}
S =\int \mathop{d^4x} \sqrt{-g}
\Bigg\{
\frac{M_{\rm pl}^2}{2}R 
-\frac{1}{2}\left(1+\frac{\sigma}{\Lambda}\right)^2g^{\mu \nu}\partial_{\mu}\phi \partial_{\nu}\phi-\frac{1}{2}f^2(\phi)g^{\mu \nu}\partial_{\mu}\sigma \partial_{\nu}\sigma-V(\phi)-\frac{1}{2}m(\phi)^2\sigma^2\,.
\Bigg\}\,.
\end{align}
Assuming that the $\sigma$ field is massive, does not spoil slow-roll inflation and does not backreact on the equations of motion for $\phi$ we have that $\sigma$ oscillates around the minimum of the effective potential given by the centrifugal force by
\begin{align}
\Delta\sigma=\Delta\sigma_r\left(\frac{a}{a_r}\right)^{-3/2}\frac{f_r}{f}\sqrt{\frac{m_{\rm eff,r}}{m_{\rm eff}}}\Big\{\cos\left(\int_{t_r}^t m_{\rm eff}\,dt\right)+O(1/\mu)\Big\}
\end{align}
where $m_{\rm eff}^2= {m^2}/{f^2}-{\ddot f}/{f}-3H{\dot f}/{f}$ and we will assume that the time derivatives of $f$ are negligible in front of $m$.  {All these conditions can be satisfied, at least momentarily, if the energy fraction of the massive field is smaller than the slow-roll parameter $\epsilon\equiv-\dot H/H^2$.} Then, the leading interacting term is given by 
\begin{align}
\frac{\Delta(\tau)}{H^2}=\frac{\ddot\sigma}{H^2\Lambda}
\end{align}
which yields
\begin{align}
C_r=\mu_r^2\frac{\Delta\sigma_r}{\Lambda} \quad\,,\,\quad\Omega_r=\frac{\mu_r}{\delta_\Omega}\,,
\end{align}
where $\mu_r\equiv\frac{m_{{\rm eff},r}}{ H_r}$,
\begin{align}
\delta_C=-\frac{3}{2}+\frac{3}{2}\delta_m-\frac{5}{2}\delta_f\quad{\rm and}\quad \delta_\Omega=\delta_m-\delta_f\,.
\end{align}
We have defined
\begin{align}
\delta_f\equiv\frac{d\ln f}{dN} \quad{\rm and}\quad \delta_m\equiv\frac{d\ln m}{dN}\,,
\end{align}
where $dN=Hdt$.

For the values used in the main text (see Tab.~\ref{table:parameters}) we have that the effective dimensionless mass and the amplitude of the field oscillation at the pivot scale respectively are
\begin{align}
\mu_*=\omega n_o\sim 15 \qquad{\rm and}\qquad \frac{\Delta\sigma_*}{\Lambda}\sim \mu_*^{-2}\sim 10^{-2}\,.
\end{align}
Thus, at the pivot scale the values are reasonable. However, we also require $|\delta_\Omega|\sim 10$ and this implies $\delta_m,\delta_f\sim 10$. This model has a growth (or decay) of the mass and/or the coefficient of the kinetic term of 4 order of magnitude per e-fold. We conclude that either there is much fine-tuning in the potential or kinetic term or the extra field will backreact in few e-folds.

\subsection{Oscillating potential}

Let us consider that the inflaton potential has an oscillating modulation of the form
\begin{align}
V=V_0(\phi)\left(1+W(\phi)\sin\left[\Omega(\phi)+\varphi\right]\right)\,.
\end{align}
In order not to spoil slow-roll, i.e. $\eta\equiv\frac{\dot\epsilon}{H\epsilon}\ll1$ we need that $W\dot \Omega^2<1$. This can be tuned by an appropriate form of $W(\phi)$. Comparing with the results of Sec.~\ref{sec:features} we find
\begin{align}
\frac{\delta\epsilon}{\epsilon}\approx\frac{W}{2}\frac{\dot\Omega}{H}\cos\left[\Omega(\tau)+\varphi\right]
\end{align}
and so
\begin{align}
\frac{\Delta(\tau)}{H^2}=\frac{\ddot\delta\epsilon}{2H^2\epsilon}
\end{align}
which yields
\begin{align}
C_r=\frac{W_r}{4}\delta_\Omega^3\Omega_r^3 \quad{\rm and}\quad \delta_C=\delta_W+3\delta_\Omega\,.
\end{align}
We have defined
\begin{align}
\delta_W\equiv\frac{d\ln W}{dN} \quad{\rm and}\quad \delta_\Omega\equiv\frac{d\ln \Omega}{dN}\,.
\end{align}

\section{Estimation of the power spectrum \label{App:spectrum}}
Here we give a brief review of the estimation for the power spectrum. The starting point is Eq.\eqref{eq:deltaP}, which using the fact that $f\propto z\,c_s^{1/2}\propto c_s^{-1/2}$  and that $s\equiv \dot c_s/Hc_s$ reads
\begin{align}\label{eq:deltaP3}
\frac{\Delta P_{\cal R}(k)}{P_{{\cal R},0}}\approx\int_{-\infty}^\infty d\ln\xi \,\left(\frac{\sin\left(2k\xi\right)}{k\xi}-\cos\left(2k\xi\right)\right)\,s(\xi)\,.
\end{align}
Now, using that
\begin{align}
c_s^2=1+BF(\ln(\tau/\tau_f))\quad{\rm where}\quad F\equiv{\rm e}^{-\beta_s^2\ln^2\left(\tau/\tau_f\right)}\quad{,}\quad B\ll1\quad{,}\quad\beta_s\gg1\,,
\end{align}
and that $s=\frac{B\beta_s}{2} \frac{dF}{dx}$, where $x\equiv-\beta_s\ln\left(\tau/\tau_f\right)$, we can write at leading order in $x/\beta_s$
\begin{align}\label{eq:deltaP4}
\frac{\Delta P_{\cal R}(k)}{P_{{\cal R},0}}&\approx\frac{B}{2}{\rm Re}\left[\int_{-\infty}^\infty dx \frac{dF}{dx}\left(1+\frac{x}{\beta_s}\right){\rm e}^{2ikc_s\tau}\right]-\frac{B}{2}{\rm Re}\left[\int_{-\infty}^\infty dx \frac{dF}{dx} {\rm e}^{2ikc_s\tau}\right]\nonumber\\&
\approx B\sqrt{\pi}\frac{k\tau_f}{\beta_s}{\rm e}^{-\frac{k^2\tau_f^2}{\beta_s^2}}\left\{\sin(2k\tau_f)+\frac{\cos(2k\tau_f)}{k\tau_f}-\frac{1}{2}\frac{\sin(2k\tau_f)}{(k\tau_f)^2}\right\}\,,
\end{align}
where used that $\xi\approx \tau$ and we expanded the mode functions as
\begin{align}
{\rm e}^{ikc_s\tau}\approx{\rm e}^{ik\tau_f}{\rm e}^{-ik\tau_fx/\beta_s}\,,
\end{align}
since $\tau\approx \tau_f\left(1-\frac{x}{\beta_s}\right)$.

\section{Estimation of the bispectrum and trispectrum \label{App:trispectrum}}

Here we briefly derive the estimate for the magnitude of the bispectrum and trispectrum. We work in the effective field theory of inflation approach \cite{Cheung:2007st,Bartolo:2013exa,Achucarro:2014msa} and expand the action up to fourth order. By picking up the terms that only involve the speed of sound $c_s$ and its derivatives at leading order in slow roll we find
\begin{align}
{S}_3=\int dt d^3x {a^3}\frac{M_{\rm pl}^2\epsilon}{H}\left\{(1-c_s^{-2})\dot{\cal R}^3+2Hs c_s^{-2}\dot{\cal R}^2{\cal R}-a^{-2}(1-c_s^{-2})\dot{\cal R}\left(\partial {\cal R}\right)^2\right\}
\end{align}
and
\begin{align}
{S}_4=\int dt d^3x& {a^4}\frac{M_{\rm pl}^2\epsilon}{4H^2}\Bigg\{-(1-c_s^{-2})\left[\dot{\cal R}^2-a^{-2}\left(\partial {\cal R}\right)^2\right]^2\nonumber\\&
-8Hs c_s^{-2}a^{-2}\left(\partial {\cal R}\right)^2\dot{\cal R}^2{\cal R}+\left(16H^2s^2-8H\dot s\right)c_s^{-2}\dot{\cal R}^2{\cal R}^2\Bigg\}\,.
\end{align}
We will use the approximation for sharp features which consists of expanding around the transition time \cite{Adshead:2011bw,Adshead:2011jq,Bartolo:2013exa,Achucarro:2014msa}.

\subsection{Bispectrum}
For the bispectrum we use the in-in formalism \cite{Maldacena:2002vr,Wang:2013eqj},
\begin{align}
\langle{\cal R}_{k_1}{\cal R}_{k_2}{\cal R}_{k_3}\rangle=-2i{\rm Re}\left[\int d\tau d^3x\langle{\cal R}_{k_1}{\cal R}_{k_2}{\cal R}_{k_3} H_{I,3}\rangle\right]
\end{align}
where $H_{I,3}=-{\cal L}_3$. As usual we use the de-Sitter mode function:
\begin{align}
{\cal R}_k=\frac{H}{\sqrt{4\epsilon c_{s,0}k^3}}\left(1+ikc_{s,0}\tau\right){\rm e}^{-ikc_s\tau}\,.
\end{align}
Now, picking up the highest contribution in terms of $\beta_s$ and evaluating the integral near the sharp feature and in the equilateral configuration ($k_1=k_2=k_3=k$ and $\mathbf{k_1}\cdot\mathbf{k_2}=-k^2/2$) we have
\begin{align}
\langle{\cal R}_{k_1}{\cal R}_{k_2}{\cal R}_{k_3}\rangle^{\rm eq}&\approx
\frac{3}{4}B\beta_s^2{\rm Im}\Bigg[\int dx \,{\rm e}^{iKc_s\tau}\Big(F\left(\frac{k\tau}{\beta_s}\right)^3-\frac{1}{2}F\beta_s^{-2}\left(\frac{k\tau}{\beta_s}\right)\left(1-i{k\tau}\right)^2\nonumber\\&
-\frac{dF}{dx}\beta_s^{-1}\left(\frac{k\tau}{\beta_s}\right)\left(1-i{k\tau}\right)\Big)\Bigg]\times \frac{P_{{\cal R},0}^2M_{\rm pl}^6}{k^6}\left(2\pi\right)^7\delta(\mathbf{k_1}+\mathbf{k_2}+\mathbf{k_3})
\\&
\approx-\frac{\sqrt{\pi}}{3}B\beta_s^2\left(\frac{K\tau_f}{2\beta_s}\right)^3{\rm e}^{-\frac{K^2\tau_f^2}{4\beta_s^2}}\sin\left(K\tau_f\right)\times \frac{P_{{\cal R},0}^2M_{\rm pl}^6}{k^6}\left(2\pi\right)^7\delta(\mathbf{k_1}+\mathbf{k_2}+\mathbf{k_3})
\end{align}
where $K=k_1+k_2+k_3$ and $P_{{\cal R},0}=\frac{H^2}{8\pi^2\epsilon c_sM_{\rm pl}^2}$. With this result we find that
\begin{align}
f_{NL}^{\rm eq}\approx-\frac{10\sqrt{\pi}}{27}B\beta_s^2\left(\frac{K\tau_f}{2\beta_s}\right)^3{\rm e}^{-\frac{K^2\tau_f^2}{4\beta_s^2}}\sin\left(K\tau_f\right)\,,
\end{align}
where we used that
\begin{align}
\langle{\cal R}_{k_1}{\cal R}_{k_2}{\cal R}_{k_3}\rangle=\frac{3}{10}f_{NL}(k_1,k_2,k_3)\frac{k_1^3+k_2^3+k_3^2}{k_1^3k_2^3k_3^3}P_{{\cal R},0}^2M_{\rm pl}^6\left(2\pi\right)^7\delta(\mathbf{k_1}+\mathbf{k_2}+\mathbf{k_3})\,.
\end{align}

\subsection{Trispectrum}
Again, for the trispectrum we will use the in-in formalism. However, this time we have the possibility of a scalar exchange \cite{Arroja:2008ga,Chen:2009bc,Wang:2013eqj}, i.e.
\begin{align}
\langle{\cal R}_{k_1}{\cal R}_{k_2}{\cal R}_{k_3}{\cal R}_{k_4}\rangle_{SE}&=\int d^3x d\tau{'}\int_{-\infty}^{\tau} d\tau{''}\langle H_{I,3}(\tau'){\cal R}_{k_1}{\cal R}_{k_2}{\cal R}_{k_3}{\cal R}_{k_4} H_{I,3}(\tau'')\rangle\nonumber\\&
-\int d^3x d\tau{'}\int_{-\infty}^{\tau'} d\tau{''}\langle H_{I,3}(\tau'')H_{I,3}(\tau'){\cal R}_{k_1}{\cal R}_{k_2}{\cal R}_{k_3}{\cal R}_{k_4} \rangle\nonumber\\&
-\int d^3x d\tau{'}\int_{-\infty}^{\tau'} d\tau{''}\langle{\cal R}_{k_1}{\cal R}_{k_2}{\cal R}_{k_3}{\cal R}_{k_4} H_{I,3}(\tau')H_{I,3}(\tau'')\rangle\,
\end{align}
and a contact interaction \cite{Chen:2009bc}, that is
\begin{align}
\langle{\cal R}_{k_1}{\cal R}_{k_2}{\cal R}_{k_3}{\cal R}_{k_4}\rangle_{CI}=-2i{\rm Re}\left[\int d\tau d^3x\langle{\cal R}_{k_1}{\cal R}_{k_2}{\cal R}_{k_3}{\cal R}_{k_4} H_{I,4}\rangle\right]\,.
\end{align}
However, a quick inspection of the scalar exchange contribution tells us that the contribution of the scalar exchange is proportional to $f_{NL}^2\approx B^2 \beta_s^4$ since, at most, there are two cubic vertex proportional to $B\beta_s^2$. As we will now see, this contribution is suppressed by a factor $B$ with respect to the leading contribution of the contact interaction which is proportional to $B\beta_s^4$ -- e.g. look at the term $\dot s$ in $S_4$ which will bring twice $\beta_s^2$ down.

Now, to simplify the computation of the interaction Hamiltonian we will assume that only the terms which are proportional to $(1-c_s^2)$ contribute to the third order Lagrangian and, thus, the third order Lagrangian is only proportional to $B$. This means that the terms in the fourth order interaction Hamiltonian that come from ${\cal L}_3$ are always squared and so proportional to $B^2$. In this way, we can neglect the terms coming from ${\cal L}_3$ and ${\cal L}_2$ and the interaction Hamiltonian is, for our purposes, given by
\begin{align}
H_{I,4}=-{\cal L}_{4}+O(B^2).
\end{align}
We again select the highest contribution in terms of $\beta_s$ and evaluate the integral near the sharp feature and in the regular tetrahedron configuration ($k_1=k_2=k_3=k_4=k$ and $|\mathbf{k_1}+\mathbf{k_3}|=|\mathbf{k_2}+\mathbf{k_4}|=k$). Then we find
\begin{align}
\langle{\cal R}_{k_1}{\cal R}_{k_2}{\cal R}_{k_3}{\cal R}_{k_4}\rangle^{\rm eq}&\approx
-\frac{1}{2}B\beta_s^4{\rm Im}\Bigg[\int dx \,{\rm e}^{iKc_s\tau}\Big(\frac{1}{8}\left(\frac{k\tau}{\beta_s}\right)^5F+2i\left(\frac{k\tau}{\beta_s}\right)^4\frac{dF}{dx}
-3\left(\frac{k\tau}{\beta_s}\right)^3\frac{d^2F}{dx^2}\Big)\Bigg]\nonumber\\&
\times \frac{P_{{\cal R},0}^3M_{\rm pl}^8}{k^9}\left(2\pi\right)^9\delta(\mathbf{k_1}+\mathbf{k_2}+\mathbf{k_3}+\mathbf{k_4})
\\&
\approx-\frac{321}{512}\sqrt{\pi}B\beta_s^4\left(\frac{K\tau_f}{2\beta_s}\right)^5{\rm e}^{-\frac{K^2\tau_f^2}{4\beta_s^2}}\sin\left(K\tau_f\right)\times \frac{P_{{\cal R},0}^3M_{\rm pl}^8}{k^9}\left(2\pi\right)^9\delta(\mathbf{k_1}+\mathbf{k_2}+\mathbf{k_3}+\mathbf{k_4})
\end{align}
where now $K=k_1+k_2+k_3+k_4$ and $P_{{\cal R},0}=\frac{H^2}{8\pi^2\epsilon c_sM_{\rm pl}^2}$.
So we have
\begin{align}
g^{\rm eq}_{NL}=-\frac{2675}{36864}\sqrt{\pi}B\beta_s^4\left(\frac{K\tau_f}{2\beta_s}\right)^5{\rm e}^{-\frac{K^2\tau_f^2}{4\beta_s^2}}\sin\left(K\tau_f\right)
\end{align}
where we used the normalization of \cite{Smith:2015uia} in order to compare with \cite{Ade:2015ava}, that is
\begin{align}
\langle{\cal R}_{k_1}{\cal R}_{k_2}{\cal R}_{k_3}{\cal R}_{k_4}\rangle=\frac{216}{100}g_{NL}(k_1,k_2,k_3)\frac{k_1^3+k_2^3+k_3^3+k_4^3}{k_1^3k_2^3k_3^3k_4^3}\times P_{{\cal R},0}^3M_{\rm pl}^8\left(2\pi\right)^9\delta(\mathbf{k_1}+\mathbf{k_2}+\mathbf{k_3}+\mathbf{k_4})\,.
\end{align}

\bibliography{biblio.bib} 

\begin{thebibliography}{79}%
\makeatletter
\providecommand \@ifxundefined [1]{%
 \@ifx{#1\undefined}
}%
\providecommand \@ifnum [1]{%
 \ifnum #1\expandafter \@firstoftwo
 \else \expandafter \@secondoftwo
 \fi
}%
\providecommand \@ifx [1]{%
 \ifx #1\expandafter \@firstoftwo
 \else \expandafter \@secondoftwo
 \fi
}%
\providecommand \natexlab [1]{#1}%
\providecommand \enquote  [1]{``#1''}%
\providecommand \bibnamefont  [1]{#1}%
\providecommand \bibfnamefont [1]{#1}%
\providecommand \citenamefont [1]{#1}%
\providecommand \href@noop [0]{\@secondoftwo}%
\providecommand \href [0]{\begingroup \@sanitize@url \@href}%
\providecommand \@href[1]{\@@startlink{#1}\@@href}%
\providecommand \@@href[1]{\endgroup#1\@@endlink}%
\providecommand \@sanitize@url [0]{\catcode `\\12\catcode `\$12\catcode
  `\&12\catcode `\#12\catcode `\^12\catcode `\_12\catcode `\%12\relax}%
\providecommand \@@startlink[1]{}%
\providecommand \@@endlink[0]{}%
\providecommand \url  [0]{\begingroup\@sanitize@url \@url }%
\providecommand \@url [1]{\endgroup\@href {#1}{\urlprefix }}%
\providecommand \urlprefix  [0]{URL }%
\providecommand \Eprint [0]{\href }%
\providecommand \doibase [0]{http://dx.doi.org/}%
\providecommand \selectlanguage [0]{\@gobble}%
\providecommand \bibinfo  [0]{\@secondoftwo}%
\providecommand \bibfield  [0]{\@secondoftwo}%
\providecommand \translation [1]{[#1]}%
\providecommand \BibitemOpen [0]{}%
\providecommand \bibitemStop [0]{}%
\providecommand \bibitemNoStop [0]{.\EOS\space}%
\providecommand \EOS [0]{\spacefactor3000\relax}%
\providecommand \BibitemShut  [1]{\csname bibitem#1\endcsname}%
\let\auto@bib@innerbib\@empty
\bibitem [{\citenamefont {Seljak}(1996)}]{Seljak:1995ve}%
  \BibitemOpen
  \bibfield  {author} {\bibinfo {author} {\bibfnamefont {Uros}\ \bibnamefont
  {Seljak}},\ }\bibfield  {title} {\enquote {\bibinfo {title} {{Gravitational
  lensing effect on cosmic microwave background anisotropies: A Power spectrum
  approach}},}\ }\href {\doibase 10.1086/177218} {\bibfield  {journal}
  {\bibinfo  {journal} {Astrophys. J.}\ }\textbf {\bibinfo {volume} {463}},\
  \bibinfo {pages} {1} (\bibinfo {year} {1996})},\ \Eprint
  {http://arxiv.org/abs/astro-ph/9505109} {arXiv:astro-ph/9505109 [astro-ph]}
  \BibitemShut {NoStop}%
\bibitem [{\citenamefont {Lewis}\ and\ \citenamefont
  {Challinor}(2006)}]{Lewis:2006fu}%
  \BibitemOpen
  \bibfield  {author} {\bibinfo {author} {\bibfnamefont {Antony}\ \bibnamefont
  {Lewis}}\ and\ \bibinfo {author} {\bibfnamefont {Anthony}\ \bibnamefont
  {Challinor}},\ }\bibfield  {title} {\enquote {\bibinfo {title} {{Weak
  gravitational lensing of the CMB}},}\ }\href {\doibase
  10.1016/j.physrep.2006.03.002} {\bibfield  {journal} {\bibinfo  {journal}
  {Phys. Rept.}\ }\textbf {\bibinfo {volume} {429}},\ \bibinfo {pages} {1--65}
  (\bibinfo {year} {2006})},\ \Eprint {http://arxiv.org/abs/astro-ph/0601594}
  {arXiv:astro-ph/0601594 [astro-ph]} \BibitemShut {NoStop}%
\bibitem [{\citenamefont {Hanson}\ \emph {et~al.}(2010)\citenamefont {Hanson},
  \citenamefont {Challinor},\ and\ \citenamefont {Lewis}}]{Hanson:2009kr}%
  \BibitemOpen
  \bibfield  {author} {\bibinfo {author} {\bibfnamefont {Duncan}\ \bibnamefont
  {Hanson}}, \bibinfo {author} {\bibfnamefont {Anthony}\ \bibnamefont
  {Challinor}}, \ and\ \bibinfo {author} {\bibfnamefont {Antony}\ \bibnamefont
  {Lewis}},\ }\bibfield  {title} {\enquote {\bibinfo {title} {{Weak lensing of
  the CMB}},}\ }\href {\doibase 10.1007/s10714-010-1036-y} {\bibfield
  {journal} {\bibinfo  {journal} {Gen. Rel. Grav.}\ }\textbf {\bibinfo {volume}
  {42}},\ \bibinfo {pages} {2197--2218} (\bibinfo {year} {2010})},\ \Eprint
  {http://arxiv.org/abs/0911.0612} {arXiv:0911.0612 [astro-ph.CO]} \BibitemShut
  {NoStop}%
\bibitem [{\citenamefont {Aghanim}\ \emph {et~al.}(2018)\citenamefont {Aghanim}
  \emph {et~al.}}]{Aghanim:2018eyx}%
  \BibitemOpen
  \bibfield  {author} {\bibinfo {author} {\bibfnamefont {N.}~\bibnamefont
  {Aghanim}} \emph {et~al.} (\bibinfo {collaboration} {Planck}),\ }\bibfield
  {title} {\enquote {\bibinfo {title} {{Planck 2018 results. VI. Cosmological
  parameters}},}\ }\href@noop {} {\  (\bibinfo {year} {2018})},\ \Eprint
  {http://arxiv.org/abs/1807.06209} {arXiv:1807.06209 [astro-ph.CO]}
  \BibitemShut {NoStop}%
\bibitem [{\citenamefont {Calabrese}\ \emph {et~al.}(2008)\citenamefont
  {Calabrese}, \citenamefont {Slosar}, \citenamefont {Melchiorri},
  \citenamefont {Smoot},\ and\ \citenamefont {Zahn}}]{Calabrese:2008rt}%
  \BibitemOpen
  \bibfield  {author} {\bibinfo {author} {\bibfnamefont {Erminia}\ \bibnamefont
  {Calabrese}}, \bibinfo {author} {\bibfnamefont {Anze}\ \bibnamefont
  {Slosar}}, \bibinfo {author} {\bibfnamefont {Alessandro}\ \bibnamefont
  {Melchiorri}}, \bibinfo {author} {\bibfnamefont {George~F.}\ \bibnamefont
  {Smoot}}, \ and\ \bibinfo {author} {\bibfnamefont {Oliver}\ \bibnamefont
  {Zahn}},\ }\bibfield  {title} {\enquote {\bibinfo {title} {{Cosmic Microwave
  Weak lensing data as a test for the dark universe}},}\ }\href {\doibase
  10.1103/PhysRevD.77.123531} {\bibfield  {journal} {\bibinfo  {journal} {Phys.
  Rev.}\ }\textbf {\bibinfo {volume} {D77}},\ \bibinfo {pages} {123531}
  (\bibinfo {year} {2008})},\ \Eprint {http://arxiv.org/abs/0803.2309}
  {arXiv:0803.2309 [astro-ph]} \BibitemShut {NoStop}%
\bibitem [{\citenamefont {Muñoz}\ \emph {et~al.}(2016)\citenamefont {Muñoz},
  \citenamefont {Grin}, \citenamefont {Dai}, \citenamefont {Kamionkowski},\
  and\ \citenamefont {Kovetz}}]{Munoz:2015fdv}%
  \BibitemOpen
  \bibfield  {author} {\bibinfo {author} {\bibfnamefont {Julian~B.}\
  \bibnamefont {Muñoz}}, \bibinfo {author} {\bibfnamefont {Daniel}\
  \bibnamefont {Grin}}, \bibinfo {author} {\bibfnamefont {Liang}\ \bibnamefont
  {Dai}}, \bibinfo {author} {\bibfnamefont {Marc}\ \bibnamefont
  {Kamionkowski}}, \ and\ \bibinfo {author} {\bibfnamefont {Ely~D.}\
  \bibnamefont {Kovetz}},\ }\bibfield  {title} {\enquote {\bibinfo {title}
  {{Search for Compensated Isocurvature Perturbations with Planck Power
  Spectra}},}\ }\href {\doibase 10.1103/PhysRevD.93.043008} {\bibfield
  {journal} {\bibinfo  {journal} {Phys. Rev.}\ }\textbf {\bibinfo {volume}
  {D93}},\ \bibinfo {pages} {043008} (\bibinfo {year} {2016})},\ \Eprint
  {http://arxiv.org/abs/1511.04441} {arXiv:1511.04441 [astro-ph.CO]}
  \BibitemShut {NoStop}%
\bibitem [{\citenamefont {Smith}\ \emph {et~al.}(2017)\citenamefont {Smith},
  \citenamefont {Muñoz}, \citenamefont {Smith}, \citenamefont {Yee},\ and\
  \citenamefont {Grin}}]{Smith:2017ndr}%
  \BibitemOpen
  \bibfield  {author} {\bibinfo {author} {\bibfnamefont {Tristan~L.}\
  \bibnamefont {Smith}}, \bibinfo {author} {\bibfnamefont {Julian~B.}\
  \bibnamefont {Muñoz}}, \bibinfo {author} {\bibfnamefont {Rhiannon}\
  \bibnamefont {Smith}}, \bibinfo {author} {\bibfnamefont {Kyle}\ \bibnamefont
  {Yee}}, \ and\ \bibinfo {author} {\bibfnamefont {Daniel}\ \bibnamefont
  {Grin}},\ }\bibfield  {title} {\enquote {\bibinfo {title} {{Baryons still
  trace dark matter: probing CMB lensing maps for hidden isocurvature}},}\
  }\href {\doibase 10.1103/PhysRevD.96.083508} {\bibfield  {journal} {\bibinfo
  {journal} {Phys. Rev.}\ }\textbf {\bibinfo {volume} {D96}},\ \bibinfo {pages}
  {083508} (\bibinfo {year} {2017})},\ \Eprint
  {http://arxiv.org/abs/1704.03461} {arXiv:1704.03461 [astro-ph.CO]}
  \BibitemShut {NoStop}%
\bibitem [{\citenamefont {Akrami}\ \emph {et~al.}(2018)\citenamefont {Akrami}
  \emph {et~al.}}]{Akrami:2018odb}%
  \BibitemOpen
  \bibfield  {author} {\bibinfo {author} {\bibfnamefont {Y.}~\bibnamefont
  {Akrami}} \emph {et~al.} (\bibinfo {collaboration} {Planck}),\ }\bibfield
  {title} {\enquote {\bibinfo {title} {{Planck 2018 results. X. Constraints on
  inflation}},}\ }\href@noop {} {\  (\bibinfo {year} {2018})},\ \Eprint
  {http://arxiv.org/abs/1807.06211} {arXiv:1807.06211 [astro-ph.CO]}
  \BibitemShut {NoStop}%
\bibitem [{\citenamefont {Hazra}\ \emph
  {et~al.}(2014{\natexlab{a}})\citenamefont {Hazra}, \citenamefont
  {Shafieloo},\ and\ \citenamefont {Souradeep}}]{Hazra:2014jwa}%
  \BibitemOpen
  \bibfield  {author} {\bibinfo {author} {\bibfnamefont {Dhiraj~Kumar}\
  \bibnamefont {Hazra}}, \bibinfo {author} {\bibfnamefont {Arman}\ \bibnamefont
  {Shafieloo}}, \ and\ \bibinfo {author} {\bibfnamefont {Tarun}\ \bibnamefont
  {Souradeep}},\ }\bibfield  {title} {\enquote {\bibinfo {title} {{Primordial
  power spectrum from Planck}},}\ }\href {\doibase
  10.1088/1475-7516/2014/11/011} {\bibfield  {journal} {\bibinfo  {journal}
  {JCAP}\ }\textbf {\bibinfo {volume} {1411}},\ \bibinfo {pages} {011}
  (\bibinfo {year} {2014}{\natexlab{a}})},\ \Eprint
  {http://arxiv.org/abs/1406.4827} {arXiv:1406.4827 [astro-ph.CO]} \BibitemShut
  {NoStop}%
\bibitem [{\citenamefont {Hazra}\ \emph {et~al.}(2019)\citenamefont {Hazra},
  \citenamefont {Shafieloo},\ and\ \citenamefont {Souradeep}}]{Hazra:2018opk}%
  \BibitemOpen
  \bibfield  {author} {\bibinfo {author} {\bibfnamefont {Dhiraj~Kumar}\
  \bibnamefont {Hazra}}, \bibinfo {author} {\bibfnamefont {Arman}\ \bibnamefont
  {Shafieloo}}, \ and\ \bibinfo {author} {\bibfnamefont {Tarun}\ \bibnamefont
  {Souradeep}},\ }\bibfield  {title} {\enquote {\bibinfo {title} {{Parameter
  discordance in Planck CMB and low-redshift measurements: projection in the
  primordial power spectrum}},}\ }\href {\doibase
  10.1088/1475-7516/2019/04/036} {\bibfield  {journal} {\bibinfo  {journal}
  {JCAP}\ }\textbf {\bibinfo {volume} {2019}},\ \bibinfo {pages} {036}
  (\bibinfo {year} {2019})},\ \Eprint {http://arxiv.org/abs/1810.08101}
  {arXiv:1810.08101 [astro-ph.CO]} \BibitemShut {NoStop}%
\bibitem [{\citenamefont {Achucarro}\ \emph {et~al.}(2014)\citenamefont
  {Achucarro}, \citenamefont {Atal}, \citenamefont {Hu}, \citenamefont
  {Ortiz},\ and\ \citenamefont {Torrado}}]{Achucarro:2014msa}%
  \BibitemOpen
  \bibfield  {author} {\bibinfo {author} {\bibfnamefont {Ana}\ \bibnamefont
  {Achucarro}}, \bibinfo {author} {\bibfnamefont {Vicente}\ \bibnamefont
  {Atal}}, \bibinfo {author} {\bibfnamefont {Bin}\ \bibnamefont {Hu}}, \bibinfo
  {author} {\bibfnamefont {Pablo}\ \bibnamefont {Ortiz}}, \ and\ \bibinfo
  {author} {\bibfnamefont {Jesus}\ \bibnamefont {Torrado}},\ }\bibfield
  {title} {\enquote {\bibinfo {title} {{Inflation with moderately sharp
  features in the speed of sound: Generalized slow roll and in-in formalism for
  power spectrum and bispectrum}},}\ }\href {\doibase
  10.1103/PhysRevD.90.023511} {\bibfield  {journal} {\bibinfo  {journal} {Phys.
  Rev.}\ }\textbf {\bibinfo {volume} {D90}},\ \bibinfo {pages} {023511}
  (\bibinfo {year} {2014})},\ \Eprint {http://arxiv.org/abs/1404.7522}
  {arXiv:1404.7522 [astro-ph.CO]} \BibitemShut {NoStop}%
\bibitem [{\citenamefont {Chen}\ and\ \citenamefont
  {Namjoo}(2014)}]{Chen:2014joa}%
  \BibitemOpen
  \bibfield  {author} {\bibinfo {author} {\bibfnamefont {Xingang}\ \bibnamefont
  {Chen}}\ and\ \bibinfo {author} {\bibfnamefont {Mohammad~Hossein}\
  \bibnamefont {Namjoo}},\ }\bibfield  {title} {\enquote {\bibinfo {title}
  {{Standard Clock in Primordial Density Perturbations and Cosmic Microwave
  Background}},}\ }\href {\doibase 10.1016/j.physletb.2014.11.002} {\bibfield
  {journal} {\bibinfo  {journal} {Phys. Lett.}\ }\textbf {\bibinfo {volume}
  {B739}},\ \bibinfo {pages} {285--292} (\bibinfo {year} {2014})},\ \Eprint
  {http://arxiv.org/abs/1404.1536} {arXiv:1404.1536 [astro-ph.CO]} \BibitemShut
  {NoStop}%
\bibitem [{\citenamefont {Hazra}\ \emph
  {et~al.}(2014{\natexlab{b}})\citenamefont {Hazra}, \citenamefont {Shafieloo},
  \citenamefont {Smoot},\ and\ \citenamefont {Starobinsky}}]{Hazra:2014goa}%
  \BibitemOpen
  \bibfield  {author} {\bibinfo {author} {\bibfnamefont {Dhiraj~Kumar}\
  \bibnamefont {Hazra}}, \bibinfo {author} {\bibfnamefont {Arman}\ \bibnamefont
  {Shafieloo}}, \bibinfo {author} {\bibfnamefont {George~F.}\ \bibnamefont
  {Smoot}}, \ and\ \bibinfo {author} {\bibfnamefont {Alexei~A.}\ \bibnamefont
  {Starobinsky}},\ }\bibfield  {title} {\enquote {\bibinfo {title} {{Wiggly
  Whipped Inflation}},}\ }\href {\doibase 10.1088/1475-7516/2014/08/048}
  {\bibfield  {journal} {\bibinfo  {journal} {JCAP}\ }\textbf {\bibinfo
  {volume} {1408}},\ \bibinfo {pages} {048} (\bibinfo {year}
  {2014}{\natexlab{b}})},\ \Eprint {http://arxiv.org/abs/1405.2012}
  {arXiv:1405.2012 [astro-ph.CO]} \BibitemShut {NoStop}%
\bibitem [{\citenamefont {Hazra}\ \emph {et~al.}(2016)\citenamefont {Hazra},
  \citenamefont {Shafieloo}, \citenamefont {Smoot},\ and\ \citenamefont
  {Starobinsky}}]{Hazra:2016fkm}%
  \BibitemOpen
  \bibfield  {author} {\bibinfo {author} {\bibfnamefont {Dhiraj~Kumar}\
  \bibnamefont {Hazra}}, \bibinfo {author} {\bibfnamefont {Arman}\ \bibnamefont
  {Shafieloo}}, \bibinfo {author} {\bibfnamefont {George~F.}\ \bibnamefont
  {Smoot}}, \ and\ \bibinfo {author} {\bibfnamefont {Alexei~A.}\ \bibnamefont
  {Starobinsky}},\ }\bibfield  {title} {\enquote {\bibinfo {title} {{Primordial
  features and Planck polarization}},}\ }\href {\doibase
  10.1088/1475-7516/2016/09/009} {\bibfield  {journal} {\bibinfo  {journal}
  {JCAP}\ }\textbf {\bibinfo {volume} {1609}},\ \bibinfo {pages} {009}
  (\bibinfo {year} {2016})},\ \Eprint {http://arxiv.org/abs/1605.02106}
  {arXiv:1605.02106 [astro-ph.CO]} \BibitemShut {NoStop}%
\bibitem [{\citenamefont {Hazra}\ \emph {et~al.}(2018)\citenamefont {Hazra},
  \citenamefont {Paoletti}, \citenamefont {Ballardini}, \citenamefont
  {Finelli}, \citenamefont {Shafieloo}, \citenamefont {Smoot},\ and\
  \citenamefont {Starobinsky}}]{Hazra:2017joc}%
  \BibitemOpen
  \bibfield  {author} {\bibinfo {author} {\bibfnamefont {Dhiraj~Kumar}\
  \bibnamefont {Hazra}}, \bibinfo {author} {\bibfnamefont {Daniela}\
  \bibnamefont {Paoletti}}, \bibinfo {author} {\bibfnamefont {Mario}\
  \bibnamefont {Ballardini}}, \bibinfo {author} {\bibfnamefont {Fabio}\
  \bibnamefont {Finelli}}, \bibinfo {author} {\bibfnamefont {Arman}\
  \bibnamefont {Shafieloo}}, \bibinfo {author} {\bibfnamefont {George~F.}\
  \bibnamefont {Smoot}}, \ and\ \bibinfo {author} {\bibfnamefont {Alexei~A.}\
  \bibnamefont {Starobinsky}},\ }\bibfield  {title} {\enquote {\bibinfo {title}
  {{Probing features in inflaton potential and reionization history with future
  CMB space observations}},}\ }\href {\doibase 10.1088/1475-7516/2018/02/017}
  {\bibfield  {journal} {\bibinfo  {journal} {JCAP}\ }\textbf {\bibinfo
  {volume} {1802}},\ \bibinfo {pages} {017} (\bibinfo {year} {2018})},\ \Eprint
  {http://arxiv.org/abs/1710.01205} {arXiv:1710.01205 [astro-ph.CO]}
  \BibitemShut {NoStop}%
\bibitem [{\citenamefont {Chluba}\ \emph {et~al.}(2015)\citenamefont {Chluba},
  \citenamefont {Hamann},\ and\ \citenamefont {Patil}}]{Chluba:2015bqa}%
  \BibitemOpen
  \bibfield  {author} {\bibinfo {author} {\bibfnamefont {Jens}\ \bibnamefont
  {Chluba}}, \bibinfo {author} {\bibfnamefont {Jan}\ \bibnamefont {Hamann}}, \
  and\ \bibinfo {author} {\bibfnamefont {Subodh~P.}\ \bibnamefont {Patil}},\
  }\bibfield  {title} {\enquote {\bibinfo {title} {{Features and New Physical
  Scales in Primordial Observables: Theory and Observation}},}\ }\href
  {\doibase 10.1142/S0218271815300232} {\bibfield  {journal} {\bibinfo
  {journal} {Int. J. Mod. Phys.}\ }\textbf {\bibinfo {volume} {D24}},\ \bibinfo
  {pages} {1530023} (\bibinfo {year} {2015})},\ \Eprint
  {http://arxiv.org/abs/1505.01834} {arXiv:1505.01834 [astro-ph.CO]}
  \BibitemShut {NoStop}%
\bibitem [{\citenamefont {Pahud}\ \emph {et~al.}(2009)\citenamefont {Pahud},
  \citenamefont {Kamionkowski},\ and\ \citenamefont {Liddle}}]{Pahud:2008ae}%
  \BibitemOpen
  \bibfield  {author} {\bibinfo {author} {\bibfnamefont {Cédric}\ \bibnamefont
  {Pahud}}, \bibinfo {author} {\bibfnamefont {Marc}\ \bibnamefont
  {Kamionkowski}}, \ and\ \bibinfo {author} {\bibfnamefont {Andrew~R.}\
  \bibnamefont {Liddle}},\ }\bibfield  {title} {\enquote {\bibinfo {title}
  {{Oscillations in the inflaton potential?}}}\ }\href {\doibase
  10.1103/PhysRevD.79.083503} {\bibfield  {journal} {\bibinfo  {journal} {Phys.
  Rev.}\ }\textbf {\bibinfo {volume} {D79}},\ \bibinfo {pages} {083503}
  (\bibinfo {year} {2009})},\ \Eprint {http://arxiv.org/abs/0807.0322}
  {arXiv:0807.0322 [astro-ph]} \BibitemShut {NoStop}%
\bibitem [{\citenamefont {Flauger}\ \emph {et~al.}(2010)\citenamefont
  {Flauger}, \citenamefont {McAllister}, \citenamefont {Pajer}, \citenamefont
  {Westphal},\ and\ \citenamefont {Xu}}]{Flauger:2009ab}%
  \BibitemOpen
  \bibfield  {author} {\bibinfo {author} {\bibfnamefont {Raphael}\ \bibnamefont
  {Flauger}}, \bibinfo {author} {\bibfnamefont {Liam}\ \bibnamefont
  {McAllister}}, \bibinfo {author} {\bibfnamefont {Enrico}\ \bibnamefont
  {Pajer}}, \bibinfo {author} {\bibfnamefont {Alexander}\ \bibnamefont
  {Westphal}}, \ and\ \bibinfo {author} {\bibfnamefont {Gang}\ \bibnamefont
  {Xu}},\ }\bibfield  {title} {\enquote {\bibinfo {title} {{Oscillations in the
  CMB from Axion Monodromy Inflation}},}\ }\href {\doibase
  10.1088/1475-7516/2010/06/009} {\bibfield  {journal} {\bibinfo  {journal}
  {JCAP}\ }\textbf {\bibinfo {volume} {1006}},\ \bibinfo {pages} {009}
  (\bibinfo {year} {2010})},\ \Eprint {http://arxiv.org/abs/0907.2916}
  {arXiv:0907.2916 [hep-th]} \BibitemShut {NoStop}%
\bibitem [{\citenamefont {Chen}(2012)}]{Chen:2011zf}%
  \BibitemOpen
  \bibfield  {author} {\bibinfo {author} {\bibfnamefont {Xingang}\ \bibnamefont
  {Chen}},\ }\bibfield  {title} {\enquote {\bibinfo {title} {{Primordial
  Features as Evidence for Inflation}},}\ }\href {\doibase
  10.1088/1475-7516/2012/01/038} {\bibfield  {journal} {\bibinfo  {journal}
  {JCAP}\ }\textbf {\bibinfo {volume} {1201}},\ \bibinfo {pages} {038}
  (\bibinfo {year} {2012})},\ \Eprint {http://arxiv.org/abs/1104.1323}
  {arXiv:1104.1323 [hep-th]} \BibitemShut {NoStop}%
\bibitem [{\citenamefont {Chen}\ \emph {et~al.}(2015)\citenamefont {Chen},
  \citenamefont {Namjoo},\ and\ \citenamefont {Wang}}]{Chen:2014cwa}%
  \BibitemOpen
  \bibfield  {author} {\bibinfo {author} {\bibfnamefont {Xingang}\ \bibnamefont
  {Chen}}, \bibinfo {author} {\bibfnamefont {Mohammad~Hossein}\ \bibnamefont
  {Namjoo}}, \ and\ \bibinfo {author} {\bibfnamefont {Yi}~\bibnamefont
  {Wang}},\ }\bibfield  {title} {\enquote {\bibinfo {title} {{Models of the
  Primordial Standard Clock}},}\ }\href {\doibase
  10.1088/1475-7516/2015/02/027} {\bibfield  {journal} {\bibinfo  {journal}
  {JCAP}\ }\textbf {\bibinfo {volume} {1502}},\ \bibinfo {pages} {027}
  (\bibinfo {year} {2015})},\ \Eprint {http://arxiv.org/abs/1411.2349}
  {arXiv:1411.2349 [astro-ph.CO]} \BibitemShut {NoStop}%
\bibitem [{\citenamefont {Chen}\ \emph
  {et~al.}(2016{\natexlab{a}})\citenamefont {Chen}, \citenamefont {Namjoo},\
  and\ \citenamefont {Wang}}]{Chen:2015lza}%
  \BibitemOpen
  \bibfield  {author} {\bibinfo {author} {\bibfnamefont {Xingang}\ \bibnamefont
  {Chen}}, \bibinfo {author} {\bibfnamefont {Mohammad~Hossein}\ \bibnamefont
  {Namjoo}}, \ and\ \bibinfo {author} {\bibfnamefont {Yi}~\bibnamefont
  {Wang}},\ }\bibfield  {title} {\enquote {\bibinfo {title} {{Quantum
  Primordial Standard Clocks}},}\ }\href {\doibase
  10.1088/1475-7516/2016/02/013} {\bibfield  {journal} {\bibinfo  {journal}
  {JCAP}\ }\textbf {\bibinfo {volume} {1602}},\ \bibinfo {pages} {013}
  (\bibinfo {year} {2016}{\natexlab{a}})},\ \Eprint
  {http://arxiv.org/abs/1509.03930} {arXiv:1509.03930 [astro-ph.CO]}
  \BibitemShut {NoStop}%
\bibitem [{\citenamefont {Gao}\ \emph {et~al.}(2013)\citenamefont {Gao},
  \citenamefont {Langlois},\ and\ \citenamefont {Mizuno}}]{Gao:2013ota}%
  \BibitemOpen
  \bibfield  {author} {\bibinfo {author} {\bibfnamefont {Xian}\ \bibnamefont
  {Gao}}, \bibinfo {author} {\bibfnamefont {David}\ \bibnamefont {Langlois}}, \
  and\ \bibinfo {author} {\bibfnamefont {Shuntaro}\ \bibnamefont {Mizuno}},\
  }\bibfield  {title} {\enquote {\bibinfo {title} {{Oscillatory features in the
  curvature power spectrum after a sudden turn of the inflationary
  trajectory}},}\ }\href {\doibase 10.1088/1475-7516/2013/10/023} {\bibfield
  {journal} {\bibinfo  {journal} {JCAP}\ }\textbf {\bibinfo {volume} {1310}},\
  \bibinfo {pages} {023} (\bibinfo {year} {2013})},\ \Eprint
  {http://arxiv.org/abs/1306.5680} {arXiv:1306.5680 [hep-th]} \BibitemShut
  {NoStop}%
\bibitem [{\citenamefont {Starobinsky}(1992)}]{Starobinsky:1992ts}%
  \BibitemOpen
  \bibfield  {author} {\bibinfo {author} {\bibfnamefont {Alexei~A.}\
  \bibnamefont {Starobinsky}},\ }\bibfield  {title} {\enquote {\bibinfo {title}
  {{Spectrum of adiabatic perturbations in the universe when there are
  singularities in the inflation potential}},}\ }\href@noop {} {\bibfield
  {journal} {\bibinfo  {journal} {JETP Lett.}\ }\textbf {\bibinfo {volume}
  {55}},\ \bibinfo {pages} {489--494} (\bibinfo {year} {1992})},\ \bibinfo
  {note} {[Pisma Zh. Eksp. Teor. Fiz.55,477(1992)]}\BibitemShut {NoStop}%
\bibitem [{\citenamefont {Kamionkowski}\ and\ \citenamefont
  {Liddle}(2000)}]{Kamionkowski:1999vp}%
  \BibitemOpen
  \bibfield  {author} {\bibinfo {author} {\bibfnamefont {Marc}\ \bibnamefont
  {Kamionkowski}}\ and\ \bibinfo {author} {\bibfnamefont {Andrew~R.}\
  \bibnamefont {Liddle}},\ }\bibfield  {title} {\enquote {\bibinfo {title}
  {{The Dearth of halo dwarf galaxies: Is there power on short scales?}}}\
  }\href {\doibase 10.1103/PhysRevLett.84.4525} {\bibfield  {journal} {\bibinfo
   {journal} {Phys. Rev. Lett.}\ }\textbf {\bibinfo {volume} {84}},\ \bibinfo
  {pages} {4525--4528} (\bibinfo {year} {2000})},\ \Eprint
  {http://arxiv.org/abs/astro-ph/9911103} {arXiv:astro-ph/9911103 [astro-ph]}
  \BibitemShut {NoStop}%
\bibitem [{\citenamefont {Gong}\ and\ \citenamefont
  {Stewart}(2001)}]{Gong:2001he}%
  \BibitemOpen
  \bibfield  {author} {\bibinfo {author} {\bibfnamefont {Jinn-Ouk}\
  \bibnamefont {Gong}}\ and\ \bibinfo {author} {\bibfnamefont {Ewan~D.}\
  \bibnamefont {Stewart}},\ }\bibfield  {title} {\enquote {\bibinfo {title}
  {{The Density perturbation power spectrum to second order corrections in the
  slow roll expansion}},}\ }\href {\doibase 10.1016/S0370-2693(01)00616-5}
  {\bibfield  {journal} {\bibinfo  {journal} {Phys. Lett.}\ }\textbf {\bibinfo
  {volume} {B510}},\ \bibinfo {pages} {1--9} (\bibinfo {year} {2001})},\
  \Eprint {http://arxiv.org/abs/astro-ph/0101225} {arXiv:astro-ph/0101225
  [astro-ph]} \BibitemShut {NoStop}%
\bibitem [{\citenamefont {Stewart}(2002)}]{Stewart:2001cd}%
  \BibitemOpen
  \bibfield  {author} {\bibinfo {author} {\bibfnamefont {Ewan~D.}\ \bibnamefont
  {Stewart}},\ }\bibfield  {title} {\enquote {\bibinfo {title} {{The Spectrum
  of density perturbations produced during inflation to leading order in a
  general slow roll approximation}},}\ }\href {\doibase
  10.1103/PhysRevD.65.103508} {\bibfield  {journal} {\bibinfo  {journal} {Phys.
  Rev.}\ }\textbf {\bibinfo {volume} {D65}},\ \bibinfo {pages} {103508}
  (\bibinfo {year} {2002})},\ \Eprint {http://arxiv.org/abs/astro-ph/0110322}
  {arXiv:astro-ph/0110322 [astro-ph]} \BibitemShut {NoStop}%
\bibitem [{\citenamefont {Adams}\ \emph {et~al.}(2001)\citenamefont {Adams},
  \citenamefont {Cresswell},\ and\ \citenamefont {Easther}}]{Adams:2001vc}%
  \BibitemOpen
  \bibfield  {author} {\bibinfo {author} {\bibfnamefont {Jennifer~A.}\
  \bibnamefont {Adams}}, \bibinfo {author} {\bibfnamefont {Bevan}\ \bibnamefont
  {Cresswell}}, \ and\ \bibinfo {author} {\bibfnamefont {Richard}\ \bibnamefont
  {Easther}},\ }\bibfield  {title} {\enquote {\bibinfo {title} {{Inflationary
  perturbations from a potential with a step}},}\ }\href {\doibase
  10.1103/PhysRevD.64.123514} {\bibfield  {journal} {\bibinfo  {journal} {Phys.
  Rev.}\ }\textbf {\bibinfo {volume} {D64}},\ \bibinfo {pages} {123514}
  (\bibinfo {year} {2001})},\ \Eprint {http://arxiv.org/abs/astro-ph/0102236}
  {arXiv:astro-ph/0102236 [astro-ph]} \BibitemShut {NoStop}%
\bibitem [{\citenamefont {Kaloper}\ and\ \citenamefont
  {Kaplinghat}(2003)}]{Kaloper:2003nv}%
  \BibitemOpen
  \bibfield  {author} {\bibinfo {author} {\bibfnamefont {Nemanja}\ \bibnamefont
  {Kaloper}}\ and\ \bibinfo {author} {\bibfnamefont {Manoj}\ \bibnamefont
  {Kaplinghat}},\ }\bibfield  {title} {\enquote {\bibinfo {title} {{Primeval
  corrections to the CMB anisotropies}},}\ }\href {\doibase
  10.1103/PhysRevD.68.123522} {\bibfield  {journal} {\bibinfo  {journal} {Phys.
  Rev.}\ }\textbf {\bibinfo {volume} {D68}},\ \bibinfo {pages} {123522}
  (\bibinfo {year} {2003})},\ \Eprint {http://arxiv.org/abs/hep-th/0307016}
  {arXiv:hep-th/0307016 [hep-th]} \BibitemShut {NoStop}%
\bibitem [{\citenamefont {Choe}\ \emph {et~al.}(2004)\citenamefont {Choe},
  \citenamefont {Gong},\ and\ \citenamefont {Stewart}}]{Choe:2004zg}%
  \BibitemOpen
  \bibfield  {author} {\bibinfo {author} {\bibfnamefont {Jeongyeol}\
  \bibnamefont {Choe}}, \bibinfo {author} {\bibfnamefont {Jinn-Ouk}\
  \bibnamefont {Gong}}, \ and\ \bibinfo {author} {\bibfnamefont {Ewan~D.}\
  \bibnamefont {Stewart}},\ }\bibfield  {title} {\enquote {\bibinfo {title}
  {{Second order general slow-roll power spectrum}},}\ }\href {\doibase
  10.1088/1475-7516/2004/07/012} {\bibfield  {journal} {\bibinfo  {journal}
  {JCAP}\ }\textbf {\bibinfo {volume} {0407}},\ \bibinfo {pages} {012}
  (\bibinfo {year} {2004})},\ \Eprint {http://arxiv.org/abs/hep-ph/0405155}
  {arXiv:hep-ph/0405155 [hep-ph]} \BibitemShut {NoStop}%
\bibitem [{\citenamefont {Dvorkin}\ and\ \citenamefont
  {Hu}(2010)}]{Dvorkin:2009ne}%
  \BibitemOpen
  \bibfield  {author} {\bibinfo {author} {\bibfnamefont {Cora}\ \bibnamefont
  {Dvorkin}}\ and\ \bibinfo {author} {\bibfnamefont {Wayne}\ \bibnamefont
  {Hu}},\ }\bibfield  {title} {\enquote {\bibinfo {title} {{Generalized Slow
  Roll for Large Power Spectrum Features}},}\ }\href {\doibase
  10.1103/PhysRevD.81.023518} {\bibfield  {journal} {\bibinfo  {journal} {Phys.
  Rev.}\ }\textbf {\bibinfo {volume} {D81}},\ \bibinfo {pages} {023518}
  (\bibinfo {year} {2010})},\ \Eprint {http://arxiv.org/abs/0910.2237}
  {arXiv:0910.2237 [astro-ph.CO]} \BibitemShut {NoStop}%
\bibitem [{\citenamefont {Achucarro}\ \emph
  {et~al.}(2011{\natexlab{a}})\citenamefont {Achucarro}, \citenamefont {Gong},
  \citenamefont {Hardeman}, \citenamefont {Palma},\ and\ \citenamefont
  {Patil}}]{Achucarro:2010da}%
  \BibitemOpen
  \bibfield  {author} {\bibinfo {author} {\bibfnamefont {Ana}\ \bibnamefont
  {Achucarro}}, \bibinfo {author} {\bibfnamefont {Jinn-Ouk}\ \bibnamefont
  {Gong}}, \bibinfo {author} {\bibfnamefont {Sjoerd}\ \bibnamefont {Hardeman}},
  \bibinfo {author} {\bibfnamefont {Gonzalo~A.}\ \bibnamefont {Palma}}, \ and\
  \bibinfo {author} {\bibfnamefont {Subodh~P.}\ \bibnamefont {Patil}},\
  }\bibfield  {title} {\enquote {\bibinfo {title} {{Features of heavy physics
  in the CMB power spectrum}},}\ }\href {\doibase
  10.1088/1475-7516/2011/01/030} {\bibfield  {journal} {\bibinfo  {journal}
  {JCAP}\ }\textbf {\bibinfo {volume} {1101}},\ \bibinfo {pages} {030}
  (\bibinfo {year} {2011}{\natexlab{a}})},\ \Eprint
  {http://arxiv.org/abs/1010.3693} {arXiv:1010.3693 [hep-ph]} \BibitemShut
  {NoStop}%
\bibitem [{\citenamefont {Adshead}\ \emph {et~al.}(2012)\citenamefont
  {Adshead}, \citenamefont {Dvorkin}, \citenamefont {Hu},\ and\ \citenamefont
  {Lim}}]{Adshead:2011jq}%
  \BibitemOpen
  \bibfield  {author} {\bibinfo {author} {\bibfnamefont {Peter}\ \bibnamefont
  {Adshead}}, \bibinfo {author} {\bibfnamefont {Cora}\ \bibnamefont {Dvorkin}},
  \bibinfo {author} {\bibfnamefont {Wayne}\ \bibnamefont {Hu}}, \ and\ \bibinfo
  {author} {\bibfnamefont {Eugene~A.}\ \bibnamefont {Lim}},\ }\bibfield
  {title} {\enquote {\bibinfo {title} {{Non-Gaussianity from Step Features in
  the Inflationary Potential}},}\ }\href {\doibase 10.1103/PhysRevD.85.023531}
  {\bibfield  {journal} {\bibinfo  {journal} {Phys. Rev.}\ }\textbf {\bibinfo
  {volume} {D85}},\ \bibinfo {pages} {023531} (\bibinfo {year} {2012})},\
  \Eprint {http://arxiv.org/abs/1110.3050} {arXiv:1110.3050 [astro-ph.CO]}
  \BibitemShut {NoStop}%
\bibitem [{\citenamefont {Shiu}\ and\ \citenamefont {Xu}(2011)}]{Shiu:2011qw}%
  \BibitemOpen
  \bibfield  {author} {\bibinfo {author} {\bibfnamefont {Gary}\ \bibnamefont
  {Shiu}}\ and\ \bibinfo {author} {\bibfnamefont {Jiajun}\ \bibnamefont {Xu}},\
  }\bibfield  {title} {\enquote {\bibinfo {title} {{Effective Field Theory and
  Decoupling in Multi-field Inflation: An Illustrative Case Study}},}\ }\href
  {\doibase 10.1103/PhysRevD.84.103509} {\bibfield  {journal} {\bibinfo
  {journal} {Phys. Rev.}\ }\textbf {\bibinfo {volume} {D84}},\ \bibinfo {pages}
  {103509} (\bibinfo {year} {2011})},\ \Eprint {http://arxiv.org/abs/1108.0981}
  {arXiv:1108.0981 [hep-th]} \BibitemShut {NoStop}%
\bibitem [{\citenamefont {Cespedes}\ \emph {et~al.}(2012)\citenamefont
  {Cespedes}, \citenamefont {Atal},\ and\ \citenamefont
  {Palma}}]{Cespedes:2012hu}%
  \BibitemOpen
  \bibfield  {author} {\bibinfo {author} {\bibfnamefont {Sebastian}\
  \bibnamefont {Cespedes}}, \bibinfo {author} {\bibfnamefont {Vicente}\
  \bibnamefont {Atal}}, \ and\ \bibinfo {author} {\bibfnamefont {Gonzalo~A.}\
  \bibnamefont {Palma}},\ }\bibfield  {title} {\enquote {\bibinfo {title} {{On
  the importance of heavy fields during inflation}},}\ }\href {\doibase
  10.1088/1475-7516/2012/05/008} {\bibfield  {journal} {\bibinfo  {journal}
  {JCAP}\ }\textbf {\bibinfo {volume} {1205}},\ \bibinfo {pages} {008}
  (\bibinfo {year} {2012})},\ \Eprint {http://arxiv.org/abs/1201.4848}
  {arXiv:1201.4848 [hep-th]} \BibitemShut {NoStop}%
\bibitem [{\citenamefont {Bartolo}\ \emph {et~al.}(2013)\citenamefont
  {Bartolo}, \citenamefont {Cannone},\ and\ \citenamefont
  {Matarrese}}]{Bartolo:2013exa}%
  \BibitemOpen
  \bibfield  {author} {\bibinfo {author} {\bibfnamefont {Nicola}\ \bibnamefont
  {Bartolo}}, \bibinfo {author} {\bibfnamefont {Dario}\ \bibnamefont
  {Cannone}}, \ and\ \bibinfo {author} {\bibfnamefont {Sabino}\ \bibnamefont
  {Matarrese}},\ }\bibfield  {title} {\enquote {\bibinfo {title} {{The
  Effective Field Theory of Inflation Models with Sharp Features}},}\ }\href
  {\doibase 10.1088/1475-7516/2013/10/038} {\bibfield  {journal} {\bibinfo
  {journal} {JCAP}\ }\textbf {\bibinfo {volume} {1310}},\ \bibinfo {pages}
  {038} (\bibinfo {year} {2013})},\ \Eprint {http://arxiv.org/abs/1307.3483}
  {arXiv:1307.3483 [astro-ph.CO]} \BibitemShut {NoStop}%
\bibitem [{\citenamefont {Easther}\ \emph {et~al.}(2005)\citenamefont
  {Easther}, \citenamefont {Kinney},\ and\ \citenamefont
  {Peiris}}]{Easther:2005yr}%
  \BibitemOpen
  \bibfield  {author} {\bibinfo {author} {\bibfnamefont {Richard}\ \bibnamefont
  {Easther}}, \bibinfo {author} {\bibfnamefont {William~H.}\ \bibnamefont
  {Kinney}}, \ and\ \bibinfo {author} {\bibfnamefont {Hiranya}\ \bibnamefont
  {Peiris}},\ }\bibfield  {title} {\enquote {\bibinfo {title} {{Boundary
  effective field theory and trans-Planckian perturbations: Astrophysical
  implications}},}\ }\href {\doibase 10.1088/1475-7516/2005/08/001} {\bibfield
  {journal} {\bibinfo  {journal} {JCAP}\ }\textbf {\bibinfo {volume} {0508}},\
  \bibinfo {pages} {001} (\bibinfo {year} {2005})},\ \Eprint
  {http://arxiv.org/abs/astro-ph/0505426} {arXiv:astro-ph/0505426 [astro-ph]}
  \BibitemShut {NoStop}%
\bibitem [{\citenamefont {Huang}\ and\ \citenamefont
  {Pi}(2018)}]{Huang:2016quc}%
  \BibitemOpen
  \bibfield  {author} {\bibinfo {author} {\bibfnamefont {Qing-Guo}\
  \bibnamefont {Huang}}\ and\ \bibinfo {author} {\bibfnamefont {Shi}\
  \bibnamefont {Pi}},\ }\bibfield  {title} {\enquote {\bibinfo {title}
  {{Power-law modulation of the scalar power spectrum from a heavy field with a
  monomial potential}},}\ }\href {\doibase 10.1088/1475-7516/2018/04/001}
  {\bibfield  {journal} {\bibinfo  {journal} {JCAP}\ }\textbf {\bibinfo
  {volume} {1804}},\ \bibinfo {pages} {001} (\bibinfo {year} {2018})},\ \Eprint
  {http://arxiv.org/abs/1610.00115} {arXiv:1610.00115 [hep-th]} \BibitemShut
  {NoStop}%
\bibitem [{\citenamefont {Domènech}\ \emph {et~al.}(2019)\citenamefont
  {Domènech}, \citenamefont {Rubio},\ and\ \citenamefont
  {Wons}}]{Domenech:2018bnf}%
  \BibitemOpen
  \bibfield  {author} {\bibinfo {author} {\bibfnamefont {Guillem}\ \bibnamefont
  {Domènech}}, \bibinfo {author} {\bibfnamefont {Javier}\ \bibnamefont
  {Rubio}}, \ and\ \bibinfo {author} {\bibfnamefont {Julius}\ \bibnamefont
  {Wons}},\ }\bibfield  {title} {\enquote {\bibinfo {title} {{Mimicking
  features in alternatives to inflation with interacting spectator fields}},}\
  }\href {\doibase 10.1016/j.physletb.2019.01.039} {\bibfield  {journal}
  {\bibinfo  {journal} {Phys. Lett.}\ }\textbf {\bibinfo {volume} {B790}},\
  \bibinfo {pages} {263--269} (\bibinfo {year} {2019})},\ \Eprint
  {http://arxiv.org/abs/1811.08224} {arXiv:1811.08224 [astro-ph.CO]}
  \BibitemShut {NoStop}%
\bibitem [{\citenamefont {Chen}\ \emph {et~al.}(2018)\citenamefont {Chen},
  \citenamefont {Loeb},\ and\ \citenamefont {Xianyu}}]{Chen:2018cgg}%
  \BibitemOpen
  \bibfield  {author} {\bibinfo {author} {\bibfnamefont {Xingang}\ \bibnamefont
  {Chen}}, \bibinfo {author} {\bibfnamefont {Abraham}\ \bibnamefont {Loeb}}, \
  and\ \bibinfo {author} {\bibfnamefont {Zhong-Zhi}\ \bibnamefont {Xianyu}},\
  }\bibfield  {title} {\enquote {\bibinfo {title} {{Unique Fingerprints of
  Alternatives to Inflation in the Primordial Power Spectrum}},}\ }\href@noop
  {} {\  (\bibinfo {year} {2018})},\ \Eprint {http://arxiv.org/abs/1809.02603}
  {arXiv:1809.02603 [astro-ph.CO]} \BibitemShut {NoStop}%
\bibitem [{\citenamefont {Carron}\ \emph {et~al.}(2017)\citenamefont {Carron},
  \citenamefont {Lewis},\ and\ \citenamefont {Challinor}}]{Carron:2017vfg}%
  \BibitemOpen
  \bibfield  {author} {\bibinfo {author} {\bibfnamefont {Julien}\ \bibnamefont
  {Carron}}, \bibinfo {author} {\bibfnamefont {Antony}\ \bibnamefont {Lewis}},
  \ and\ \bibinfo {author} {\bibfnamefont {Anthony}\ \bibnamefont
  {Challinor}},\ }\bibfield  {title} {\enquote {\bibinfo {title} {{Internal
  delensing of Planck CMB temperature and polarization}},}\ }\href {\doibase
  10.1088/1475-7516/2017/05/035} {\bibfield  {journal} {\bibinfo  {journal}
  {JCAP}\ }\textbf {\bibinfo {volume} {1705}},\ \bibinfo {pages} {035}
  (\bibinfo {year} {2017})},\ \Eprint {http://arxiv.org/abs/1701.01712}
  {arXiv:1701.01712 [astro-ph.CO]} \BibitemShut {NoStop}%
\bibitem [{\citenamefont {Lesgourgues}(2011)}]{Lesgourgues:2011re}%
  \BibitemOpen
  \bibfield  {author} {\bibinfo {author} {\bibfnamefont {Julien}\ \bibnamefont
  {Lesgourgues}},\ }\bibfield  {title} {\enquote {\bibinfo {title} {{The Cosmic
  Linear Anisotropy Solving System (CLASS) I: Overview}},}\ }\href@noop {} {\
  (\bibinfo {year} {2011})},\ \Eprint {http://arxiv.org/abs/1104.2932}
  {arXiv:1104.2932 [astro-ph.IM]} \BibitemShut {NoStop}%
\bibitem [{\citenamefont {Blas}\ \emph {et~al.}(2011)\citenamefont {Blas},
  \citenamefont {Lesgourgues},\ and\ \citenamefont {Tram}}]{Blas:2011rf}%
  \BibitemOpen
  \bibfield  {author} {\bibinfo {author} {\bibfnamefont {Diego}\ \bibnamefont
  {Blas}}, \bibinfo {author} {\bibfnamefont {Julien}\ \bibnamefont
  {Lesgourgues}}, \ and\ \bibinfo {author} {\bibfnamefont {Thomas}\
  \bibnamefont {Tram}},\ }\bibfield  {title} {\enquote {\bibinfo {title} {{The
  Cosmic Linear Anisotropy Solving System (CLASS) II: Approximation
  schemes}},}\ }\href {\doibase 10.1088/1475-7516/2011/07/034} {\bibfield
  {journal} {\bibinfo  {journal} {JCAP}\ }\textbf {\bibinfo {volume} {1107}},\
  \bibinfo {pages} {034} (\bibinfo {year} {2011})},\ \Eprint
  {http://arxiv.org/abs/1104.2933} {arXiv:1104.2933 [astro-ph.CO]} \BibitemShut
  {NoStop}%
\bibitem [{\citenamefont {Hu}(1995)}]{Hu:1995em}%
  \BibitemOpen
  \bibfield  {author} {\bibinfo {author} {\bibfnamefont {Wayne~T.}\
  \bibnamefont {Hu}},\ }\emph {\bibinfo {title} {{Wandering in the Background:
  A CMB Explorer}}},\ \href@noop {} {Ph.D. thesis},\ \bibinfo  {school} {UC,
  Berkeley} (\bibinfo {year} {1995}),\ \Eprint
  {http://arxiv.org/abs/astro-ph/9508126} {arXiv:astro-ph/9508126 [astro-ph]}
  \BibitemShut {NoStop}%
\bibitem [{\citenamefont {Silverstein}\ and\ \citenamefont
  {Westphal}(2008)}]{Silverstein:2008sg}%
  \BibitemOpen
  \bibfield  {author} {\bibinfo {author} {\bibfnamefont {Eva}\ \bibnamefont
  {Silverstein}}\ and\ \bibinfo {author} {\bibfnamefont {Alexander}\
  \bibnamefont {Westphal}},\ }\bibfield  {title} {\enquote {\bibinfo {title}
  {{Monodromy in the CMB: Gravity Waves and String Inflation}},}\ }\href
  {\doibase 10.1103/PhysRevD.78.106003} {\bibfield  {journal} {\bibinfo
  {journal} {Phys. Rev.}\ }\textbf {\bibinfo {volume} {D78}},\ \bibinfo {pages}
  {106003} (\bibinfo {year} {2008})},\ \Eprint {http://arxiv.org/abs/0803.3085}
  {arXiv:0803.3085 [hep-th]} \BibitemShut {NoStop}%
\bibitem [{\citenamefont {Achucarro}\ \emph
  {et~al.}(2011{\natexlab{b}})\citenamefont {Achucarro}, \citenamefont {Gong},
  \citenamefont {Hardeman}, \citenamefont {Palma},\ and\ \citenamefont
  {Patil}}]{Achucarro:2010jv}%
  \BibitemOpen
  \bibfield  {author} {\bibinfo {author} {\bibfnamefont {Ana}\ \bibnamefont
  {Achucarro}}, \bibinfo {author} {\bibfnamefont {Jinn-Ouk}\ \bibnamefont
  {Gong}}, \bibinfo {author} {\bibfnamefont {Sjoerd}\ \bibnamefont {Hardeman}},
  \bibinfo {author} {\bibfnamefont {Gonzalo~A.}\ \bibnamefont {Palma}}, \ and\
  \bibinfo {author} {\bibfnamefont {Subodh~P.}\ \bibnamefont {Patil}},\
  }\bibfield  {title} {\enquote {\bibinfo {title} {{Mass hierarchies and
  non-decoupling in multi-scalar field dynamics}},}\ }\href {\doibase
  10.1103/PhysRevD.84.043502} {\bibfield  {journal} {\bibinfo  {journal} {Phys.
  Rev.}\ }\textbf {\bibinfo {volume} {D84}},\ \bibinfo {pages} {043502}
  (\bibinfo {year} {2011}{\natexlab{b}})},\ \Eprint
  {http://arxiv.org/abs/1005.3848} {arXiv:1005.3848 [hep-th]} \BibitemShut
  {NoStop}%
\bibitem [{\citenamefont {Achucarro}\ \emph {et~al.}(2012)\citenamefont
  {Achucarro}, \citenamefont {Gong}, \citenamefont {Hardeman}, \citenamefont
  {Palma},\ and\ \citenamefont {Patil}}]{Achucarro:2012sm}%
  \BibitemOpen
  \bibfield  {author} {\bibinfo {author} {\bibfnamefont {Ana}\ \bibnamefont
  {Achucarro}}, \bibinfo {author} {\bibfnamefont {Jinn-Ouk}\ \bibnamefont
  {Gong}}, \bibinfo {author} {\bibfnamefont {Sjoerd}\ \bibnamefont {Hardeman}},
  \bibinfo {author} {\bibfnamefont {Gonzalo~A.}\ \bibnamefont {Palma}}, \ and\
  \bibinfo {author} {\bibfnamefont {Subodh~P.}\ \bibnamefont {Patil}},\
  }\bibfield  {title} {\enquote {\bibinfo {title} {{Effective theories of
  single field inflation when heavy fields matter}},}\ }\href {\doibase
  10.1007/JHEP05(2012)066} {\bibfield  {journal} {\bibinfo  {journal} {JHEP}\
  }\textbf {\bibinfo {volume} {05}},\ \bibinfo {pages} {066} (\bibinfo {year}
  {2012})},\ \Eprint {http://arxiv.org/abs/1201.6342} {arXiv:1201.6342
  [hep-th]} \BibitemShut {NoStop}%
\bibitem [{\citenamefont {Palma}(2015)}]{Palma:2014hra}%
  \BibitemOpen
  \bibfield  {author} {\bibinfo {author} {\bibfnamefont {Gonzalo~A.}\
  \bibnamefont {Palma}},\ }\bibfield  {title} {\enquote {\bibinfo {title}
  {{Untangling features in the primordial spectra}},}\ }\href {\doibase
  10.1088/1475-7516/2015/04/035} {\bibfield  {journal} {\bibinfo  {journal}
  {JCAP}\ }\textbf {\bibinfo {volume} {1504}},\ \bibinfo {pages} {035}
  (\bibinfo {year} {2015})},\ \Eprint {http://arxiv.org/abs/1412.5615}
  {arXiv:1412.5615 [hep-th]} \BibitemShut {NoStop}%
\bibitem [{\citenamefont {Kodama}\ and\ \citenamefont
  {Sasaki}(1984)}]{Kodama:1985bj}%
  \BibitemOpen
  \bibfield  {author} {\bibinfo {author} {\bibfnamefont {Hideo}\ \bibnamefont
  {Kodama}}\ and\ \bibinfo {author} {\bibfnamefont {Misao}\ \bibnamefont
  {Sasaki}},\ }\bibfield  {title} {\enquote {\bibinfo {title} {{Cosmological
  Perturbation Theory}},}\ }\href {\doibase 10.1143/PTPS.78.1} {\bibfield
  {journal} {\bibinfo  {journal} {Prog. Theor. Phys. Suppl.}\ }\textbf
  {\bibinfo {volume} {78}},\ \bibinfo {pages} {1--166} (\bibinfo {year}
  {1984})}\BibitemShut {NoStop}%
\bibitem [{\citenamefont {Mukhanov}\ \emph {et~al.}(1992)\citenamefont
  {Mukhanov}, \citenamefont {Feldman},\ and\ \citenamefont
  {Brandenberger}}]{Mukhanov:1990me}%
  \BibitemOpen
  \bibfield  {author} {\bibinfo {author} {\bibfnamefont {Viatcheslav~F.}\
  \bibnamefont {Mukhanov}}, \bibinfo {author} {\bibfnamefont {H.~A.}\
  \bibnamefont {Feldman}}, \ and\ \bibinfo {author} {\bibfnamefont {Robert~H.}\
  \bibnamefont {Brandenberger}},\ }\bibfield  {title} {\enquote {\bibinfo
  {title} {{Theory of cosmological perturbations. Part 1. Classical
  perturbations. Part 2. Quantum theory of perturbations. Part 3.
  Extensions}},}\ }\href {\doibase 10.1016/0370-1573(92)90044-Z} {\bibfield
  {journal} {\bibinfo  {journal} {Phys. Rept.}\ }\textbf {\bibinfo {volume}
  {215}},\ \bibinfo {pages} {203--333} (\bibinfo {year} {1992})}\BibitemShut
  {NoStop}%
\bibitem [{\citenamefont {Joy}\ \emph {et~al.}(2005)\citenamefont {Joy},
  \citenamefont {Stewart}, \citenamefont {Gong},\ and\ \citenamefont
  {Lee}}]{Joy:2005ep}%
  \BibitemOpen
  \bibfield  {author} {\bibinfo {author} {\bibfnamefont {Minu}\ \bibnamefont
  {Joy}}, \bibinfo {author} {\bibfnamefont {Ewan~D.}\ \bibnamefont {Stewart}},
  \bibinfo {author} {\bibfnamefont {Jinn-Ouk}\ \bibnamefont {Gong}}, \ and\
  \bibinfo {author} {\bibfnamefont {Hyun-Chul}\ \bibnamefont {Lee}},\
  }\bibfield  {title} {\enquote {\bibinfo {title} {{From the spectrum to
  inflation: An Inverse formula for the general slow-roll spectrum}},}\ }\href
  {\doibase 10.1088/1475-7516/2005/04/012} {\bibfield  {journal} {\bibinfo
  {journal} {JCAP}\ }\textbf {\bibinfo {volume} {0504}},\ \bibinfo {pages}
  {012} (\bibinfo {year} {2005})},\ \Eprint
  {http://arxiv.org/abs/astro-ph/0501659} {arXiv:astro-ph/0501659 [astro-ph]}
  \BibitemShut {NoStop}%
\bibitem [{\citenamefont {Gong}(2005)}]{Gong:2005jr}%
  \BibitemOpen
  \bibfield  {author} {\bibinfo {author} {\bibfnamefont {Jinn-Ouk}\
  \bibnamefont {Gong}},\ }\bibfield  {title} {\enquote {\bibinfo {title}
  {{Breaking scale invariance from a singular inflaton potential}},}\ }\href
  {\doibase 10.1088/1475-7516/2005/07/015} {\bibfield  {journal} {\bibinfo
  {journal} {JCAP}\ }\textbf {\bibinfo {volume} {0507}},\ \bibinfo {pages}
  {015} (\bibinfo {year} {2005})},\ \Eprint
  {http://arxiv.org/abs/astro-ph/0504383} {arXiv:astro-ph/0504383 [astro-ph]}
  \BibitemShut {NoStop}%
\bibitem [{\citenamefont {Durakovic}\ \emph {et~al.}(2019)\citenamefont
  {Durakovic}, \citenamefont {Hunt}, \citenamefont {Patil},\ and\ \citenamefont
  {Sarkar}}]{Durakovic:2019kqq}%
  \BibitemOpen
  \bibfield  {author} {\bibinfo {author} {\bibfnamefont {Amel}\ \bibnamefont
  {Durakovic}}, \bibinfo {author} {\bibfnamefont {Paul}\ \bibnamefont {Hunt}},
  \bibinfo {author} {\bibfnamefont {Subodh~P.}\ \bibnamefont {Patil}}, \ and\
  \bibinfo {author} {\bibfnamefont {Subir}\ \bibnamefont {Sarkar}},\ }\bibfield
   {title} {\enquote {\bibinfo {title} {{Reconstructing the EFT of Inflation
  from Cosmological Data}},}\ }\href@noop {} {\  (\bibinfo {year} {2019})},\
  \Eprint {http://arxiv.org/abs/1904.00991} {arXiv:1904.00991 [astro-ph.CO]}
  \BibitemShut {NoStop}%
\bibitem [{\citenamefont {Hu}(2011)}]{Hu:2011vr}%
  \BibitemOpen
  \bibfield  {author} {\bibinfo {author} {\bibfnamefont {Wayne}\ \bibnamefont
  {Hu}},\ }\bibfield  {title} {\enquote {\bibinfo {title} {{Generalized Slow
  Roll for Non-Canonical Kinetic Terms}},}\ }\href {\doibase
  10.1103/PhysRevD.84.027303} {\bibfield  {journal} {\bibinfo  {journal} {Phys.
  Rev.}\ }\textbf {\bibinfo {volume} {D84}},\ \bibinfo {pages} {027303}
  (\bibinfo {year} {2011})},\ \Eprint {http://arxiv.org/abs/1104.4500}
  {arXiv:1104.4500 [astro-ph.CO]} \BibitemShut {NoStop}%
\bibitem [{\citenamefont {Starobinsky}(1985)}]{Starobinsky:1986fxa}%
  \BibitemOpen
  \bibfield  {author} {\bibinfo {author} {\bibfnamefont {Alexei~A.}\
  \bibnamefont {Starobinsky}},\ }\bibfield  {title} {\enquote {\bibinfo {title}
  {{Multicomponent de Sitter (Inflationary) Stages and the Generation of
  Perturbations}},}\ }\href@noop {} {\bibfield  {journal} {\bibinfo  {journal}
  {JETP Lett.}\ }\textbf {\bibinfo {volume} {42}},\ \bibinfo {pages} {152--155}
  (\bibinfo {year} {1985})},\ \bibinfo {note} {[Pisma Zh. Eksp. Teor.
  Fiz.42,124(1985)]}\BibitemShut {NoStop}%
\bibitem [{\citenamefont {Salopek}\ and\ \citenamefont
  {Bond}(1990)}]{Salopek:1990jq}%
  \BibitemOpen
  \bibfield  {author} {\bibinfo {author} {\bibfnamefont {D.~S.}\ \bibnamefont
  {Salopek}}\ and\ \bibinfo {author} {\bibfnamefont {J.~R.}\ \bibnamefont
  {Bond}},\ }\bibfield  {title} {\enquote {\bibinfo {title} {{Nonlinear
  evolution of long wavelength metric fluctuations in inflationary models}},}\
  }\href {\doibase 10.1103/PhysRevD.42.3936} {\bibfield  {journal} {\bibinfo
  {journal} {Phys. Rev.}\ }\textbf {\bibinfo {volume} {D42}},\ \bibinfo {pages}
  {3936--3962} (\bibinfo {year} {1990})}\BibitemShut {NoStop}%
\bibitem [{\citenamefont {Sasaki}\ and\ \citenamefont
  {Stewart}(1996)}]{Sasaki:1995aw}%
  \BibitemOpen
  \bibfield  {author} {\bibinfo {author} {\bibfnamefont {Misao}\ \bibnamefont
  {Sasaki}}\ and\ \bibinfo {author} {\bibfnamefont {Ewan~D.}\ \bibnamefont
  {Stewart}},\ }\bibfield  {title} {\enquote {\bibinfo {title} {{A General
  analytic formula for the spectral index of the density perturbations produced
  during inflation}},}\ }\href {\doibase 10.1143/PTP.95.71} {\bibfield
  {journal} {\bibinfo  {journal} {Prog. Theor. Phys.}\ }\textbf {\bibinfo
  {volume} {95}},\ \bibinfo {pages} {71--78} (\bibinfo {year} {1996})},\
  \Eprint {http://arxiv.org/abs/astro-ph/9507001} {arXiv:astro-ph/9507001
  [astro-ph]} \BibitemShut {NoStop}%
\bibitem [{\citenamefont {Babichev}\ \emph {et~al.}(2008)\citenamefont
  {Babichev}, \citenamefont {Mukhanov},\ and\ \citenamefont
  {Vikman}}]{Babichev:2007dw}%
  \BibitemOpen
  \bibfield  {author} {\bibinfo {author} {\bibfnamefont {Eugeny}\ \bibnamefont
  {Babichev}}, \bibinfo {author} {\bibfnamefont {Viatcheslav}\ \bibnamefont
  {Mukhanov}}, \ and\ \bibinfo {author} {\bibfnamefont {Alexander}\
  \bibnamefont {Vikman}},\ }\bibfield  {title} {\enquote {\bibinfo {title}
  {{k-Essence, superluminal propagation, causality and emergent geometry}},}\
  }\href {\doibase 10.1088/1126-6708/2008/02/101} {\bibfield  {journal}
  {\bibinfo  {journal} {JHEP}\ }\textbf {\bibinfo {volume} {02}},\ \bibinfo
  {pages} {101} (\bibinfo {year} {2008})},\ \Eprint
  {http://arxiv.org/abs/0708.0561} {arXiv:0708.0561 [hep-th]} \BibitemShut
  {NoStop}%
\bibitem [{\citenamefont {Adams}\ \emph {et~al.}(2006)\citenamefont {Adams},
  \citenamefont {Arkani-Hamed}, \citenamefont {Dubovsky}, \citenamefont
  {Nicolis},\ and\ \citenamefont {Rattazzi}}]{Adams:2006sv}%
  \BibitemOpen
  \bibfield  {author} {\bibinfo {author} {\bibfnamefont {Allan}\ \bibnamefont
  {Adams}}, \bibinfo {author} {\bibfnamefont {Nima}\ \bibnamefont
  {Arkani-Hamed}}, \bibinfo {author} {\bibfnamefont {Sergei}\ \bibnamefont
  {Dubovsky}}, \bibinfo {author} {\bibfnamefont {Alberto}\ \bibnamefont
  {Nicolis}}, \ and\ \bibinfo {author} {\bibfnamefont {Riccardo}\ \bibnamefont
  {Rattazzi}},\ }\bibfield  {title} {\enquote {\bibinfo {title} {{Causality,
  analyticity and an IR obstruction to UV completion}},}\ }\href {\doibase
  10.1088/1126-6708/2006/10/014} {\bibfield  {journal} {\bibinfo  {journal}
  {JHEP}\ }\textbf {\bibinfo {volume} {10}},\ \bibinfo {pages} {014} (\bibinfo
  {year} {2006})},\ \Eprint {http://arxiv.org/abs/hep-th/0602178}
  {arXiv:hep-th/0602178 [hep-th]} \BibitemShut {NoStop}%
\bibitem [{\citenamefont {Mukhanov}\ and\ \citenamefont
  {Vikman}(2006)}]{Mukhanov:2005bu}%
  \BibitemOpen
  \bibfield  {author} {\bibinfo {author} {\bibfnamefont {Viatcheslav~F.}\
  \bibnamefont {Mukhanov}}\ and\ \bibinfo {author} {\bibfnamefont {Alexander}\
  \bibnamefont {Vikman}},\ }\bibfield  {title} {\enquote {\bibinfo {title}
  {{Enhancing the tensor-to-scalar ratio in simple inflation}},}\ }\href
  {\doibase 10.1088/1475-7516/2006/02/004} {\bibfield  {journal} {\bibinfo
  {journal} {JCAP}\ }\textbf {\bibinfo {volume} {0602}},\ \bibinfo {pages}
  {004} (\bibinfo {year} {2006})},\ \Eprint
  {http://arxiv.org/abs/astro-ph/0512066} {arXiv:astro-ph/0512066 [astro-ph]}
  \BibitemShut {NoStop}%
\bibitem [{\citenamefont {Ade}\ \emph {et~al.}(2016)\citenamefont {Ade} \emph
  {et~al.}}]{Ade:2015ava}%
  \BibitemOpen
  \bibfield  {author} {\bibinfo {author} {\bibfnamefont {P.~A.~R.}\
  \bibnamefont {Ade}} \emph {et~al.} (\bibinfo {collaboration} {Planck}),\
  }\bibfield  {title} {\enquote {\bibinfo {title} {{Planck 2015 results. XVII.
  Constraints on primordial non-Gaussianity}},}\ }\href {\doibase
  10.1051/0004-6361/201525836} {\bibfield  {journal} {\bibinfo  {journal}
  {Astron. Astrophys.}\ }\textbf {\bibinfo {volume} {594}},\ \bibinfo {pages}
  {A17} (\bibinfo {year} {2016})},\ \Eprint {http://arxiv.org/abs/1502.01592}
  {arXiv:1502.01592 [astro-ph.CO]} \BibitemShut {NoStop}%
\bibitem [{\citenamefont {Achúcarro}\ \emph {et~al.}(2013)\citenamefont
  {Achúcarro}, \citenamefont {Gong}, \citenamefont {Palma},\ and\
  \citenamefont {Patil}}]{Achucarro:2012fd}%
  \BibitemOpen
  \bibfield  {author} {\bibinfo {author} {\bibfnamefont {Ana}\ \bibnamefont
  {Achúcarro}}, \bibinfo {author} {\bibfnamefont {Jinn-Ouk}\ \bibnamefont
  {Gong}}, \bibinfo {author} {\bibfnamefont {Gonzalo~A.}\ \bibnamefont
  {Palma}}, \ and\ \bibinfo {author} {\bibfnamefont {Subodh~P.}\ \bibnamefont
  {Patil}},\ }\bibfield  {title} {\enquote {\bibinfo {title} {{Correlating
  features in the primordial spectra}},}\ }\href {\doibase
  10.1103/PhysRevD.87.121301} {\bibfield  {journal} {\bibinfo  {journal} {Phys.
  Rev.}\ }\textbf {\bibinfo {volume} {D87}},\ \bibinfo {pages} {121301}
  (\bibinfo {year} {2013})},\ \Eprint {http://arxiv.org/abs/1211.5619}
  {arXiv:1211.5619 [astro-ph.CO]} \BibitemShut {NoStop}%
\bibitem [{\citenamefont {Mooij}\ \emph {et~al.}(2016)\citenamefont {Mooij},
  \citenamefont {Palma}, \citenamefont {Panotopoulos},\ and\ \citenamefont
  {Soto}}]{Mooij:2016dsi}%
  \BibitemOpen
  \bibfield  {author} {\bibinfo {author} {\bibfnamefont {Sander}\ \bibnamefont
  {Mooij}}, \bibinfo {author} {\bibfnamefont {Gonzalo~A.}\ \bibnamefont
  {Palma}}, \bibinfo {author} {\bibfnamefont {Grigoris}\ \bibnamefont
  {Panotopoulos}}, \ and\ \bibinfo {author} {\bibfnamefont {Alex}\ \bibnamefont
  {Soto}},\ }\bibfield  {title} {\enquote {\bibinfo {title} {{Consistency
  relations for sharp inflationary non-Gaussian features}},}\ }\href {\doibase
  10.1088/1475-7516/2016/09/004} {\bibfield  {journal} {\bibinfo  {journal}
  {JCAP}\ }\textbf {\bibinfo {volume} {1609}},\ \bibinfo {pages} {004}
  (\bibinfo {year} {2016})},\ \Eprint {http://arxiv.org/abs/1604.03533}
  {arXiv:1604.03533 [astro-ph.CO]} \BibitemShut {NoStop}%
\bibitem [{\citenamefont {Nakashima}\ \emph {et~al.}(2011)\citenamefont
  {Nakashima}, \citenamefont {Saito}, \citenamefont {Takamizu},\ and\
  \citenamefont {Yokoyama}}]{Nakashima:2010sa}%
  \BibitemOpen
  \bibfield  {author} {\bibinfo {author} {\bibfnamefont {Masahiro}\
  \bibnamefont {Nakashima}}, \bibinfo {author} {\bibfnamefont {Ryo}\
  \bibnamefont {Saito}}, \bibinfo {author} {\bibfnamefont {Yu-ichi}\
  \bibnamefont {Takamizu}}, \ and\ \bibinfo {author} {\bibfnamefont {Jun'ichi}\
  \bibnamefont {Yokoyama}},\ }\bibfield  {title} {\enquote {\bibinfo {title}
  {{The effect of varying sound velocity on primordial curvature
  perturbations}},}\ }\href {\doibase 10.1143/PTP.125.1035} {\bibfield
  {journal} {\bibinfo  {journal} {Prog. Theor. Phys.}\ }\textbf {\bibinfo
  {volume} {125}},\ \bibinfo {pages} {1035--1052} (\bibinfo {year} {2011})},\
  \Eprint {http://arxiv.org/abs/1009.4394} {arXiv:1009.4394 [astro-ph.CO]}
  \BibitemShut {NoStop}%
\bibitem [{\citenamefont {Zeng}\ \emph {et~al.}(2019)\citenamefont {Zeng},
  \citenamefont {Kovetz}, \citenamefont {Chen}, \citenamefont {Gong},
  \citenamefont {Muñoz},\ and\ \citenamefont {Kamionkowski}}]{zeng:2018ufm}%
  \BibitemOpen
  \bibfield  {author} {\bibinfo {author} {\bibfnamefont {Chenxiao}\
  \bibnamefont {Zeng}}, \bibinfo {author} {\bibfnamefont {Ely~D.}\ \bibnamefont
  {Kovetz}}, \bibinfo {author} {\bibfnamefont {Xuelei}\ \bibnamefont {Chen}},
  \bibinfo {author} {\bibfnamefont {Yan}\ \bibnamefont {Gong}}, \bibinfo
  {author} {\bibfnamefont {Julian~B.}\ \bibnamefont {Muñoz}}, \ and\ \bibinfo
  {author} {\bibfnamefont {Marc}\ \bibnamefont {Kamionkowski}},\ }\bibfield
  {title} {\enquote {\bibinfo {title} {{Searching for Oscillations in the
  Primordial Power Spectrum with CMB and LSS Data}},}\ }\href {\doibase
  10.1103/PhysRevD.99.043517} {\bibfield  {journal} {\bibinfo  {journal} {Phys.
  Rev.}\ }\textbf {\bibinfo {volume} {D99}},\ \bibinfo {pages} {043517}
  (\bibinfo {year} {2019})},\ \Eprint {http://arxiv.org/abs/1812.05105}
  {arXiv:1812.05105 [astro-ph.CO]} \BibitemShut {NoStop}%
\bibitem [{\citenamefont {Eisenstein}\ and\ \citenamefont
  {Hu}(1998)}]{eisenstein:1997ik}%
  \BibitemOpen
  \bibfield  {author} {\bibinfo {author} {\bibfnamefont {Daniel~J.}\
  \bibnamefont {Eisenstein}}\ and\ \bibinfo {author} {\bibfnamefont {Wayne}\
  \bibnamefont {Hu}},\ }\bibfield  {title} {\enquote {\bibinfo {title}
  {{Baryonic features in the matter transfer function}},}\ }\href {\doibase
  10.1086/305424} {\bibfield  {journal} {\bibinfo  {journal} {Astrophys. J.}\
  }\textbf {\bibinfo {volume} {496}},\ \bibinfo {pages} {605} (\bibinfo {year}
  {1998})},\ \Eprint {http://arxiv.org/abs/astro-ph/9709112}
  {arXiv:astro-ph/9709112 [astro-ph]} \BibitemShut {NoStop}%
\bibitem [{\citenamefont {Gong}\ \emph {et~al.}(2014)\citenamefont {Gong},
  \citenamefont {Schalm},\ and\ \citenamefont {Shiu}}]{Gong:2014spa}%
  \BibitemOpen
  \bibfield  {author} {\bibinfo {author} {\bibfnamefont {Jinn-Ouk}\
  \bibnamefont {Gong}}, \bibinfo {author} {\bibfnamefont {Koenraad}\
  \bibnamefont {Schalm}}, \ and\ \bibinfo {author} {\bibfnamefont {Gary}\
  \bibnamefont {Shiu}},\ }\bibfield  {title} {\enquote {\bibinfo {title}
  {{Correlating correlation functions of primordial perturbations}},}\ }\href
  {\doibase 10.1103/PhysRevD.89.063540} {\bibfield  {journal} {\bibinfo
  {journal} {Phys. Rev.}\ }\textbf {\bibinfo {volume} {D89}},\ \bibinfo {pages}
  {063540} (\bibinfo {year} {2014})},\ \Eprint {http://arxiv.org/abs/1401.4402}
  {arXiv:1401.4402 [astro-ph.CO]} \BibitemShut {NoStop}%
\bibitem [{\citenamefont {Chen}\ \emph
  {et~al.}(2016{\natexlab{b}})\citenamefont {Chen}, \citenamefont {Dvorkin},
  \citenamefont {Huang}, \citenamefont {Namjoo},\ and\ \citenamefont
  {Verde}}]{Chen:2016vvw}%
  \BibitemOpen
  \bibfield  {author} {\bibinfo {author} {\bibfnamefont {Xingang}\ \bibnamefont
  {Chen}}, \bibinfo {author} {\bibfnamefont {Cora}\ \bibnamefont {Dvorkin}},
  \bibinfo {author} {\bibfnamefont {Zhiqi}\ \bibnamefont {Huang}}, \bibinfo
  {author} {\bibfnamefont {Mohammad~Hossein}\ \bibnamefont {Namjoo}}, \ and\
  \bibinfo {author} {\bibfnamefont {Licia}\ \bibnamefont {Verde}},\ }\bibfield
  {title} {\enquote {\bibinfo {title} {{The Future of Primordial Features with
  Large-Scale Structure Surveys}},}\ }\href {\doibase
  10.1088/1475-7516/2016/11/014} {\bibfield  {journal} {\bibinfo  {journal}
  {JCAP}\ }\textbf {\bibinfo {volume} {1611}},\ \bibinfo {pages} {014}
  (\bibinfo {year} {2016}{\natexlab{b}})},\ \Eprint
  {http://arxiv.org/abs/1605.09365} {arXiv:1605.09365 [astro-ph.CO]}
  \BibitemShut {NoStop}%
\bibitem [{\citenamefont {Finelli}\ \emph {et~al.}(2018)\citenamefont {Finelli}
  \emph {et~al.}}]{Finelli:2016cyd}%
  \BibitemOpen
  \bibfield  {author} {\bibinfo {author} {\bibfnamefont {Fabio}\ \bibnamefont
  {Finelli}} \emph {et~al.} (\bibinfo {collaboration} {CORE}),\ }\bibfield
  {title} {\enquote {\bibinfo {title} {{Exploring cosmic origins with CORE:
  Inflation}},}\ }\href {\doibase 10.1088/1475-7516/2018/04/016} {\bibfield
  {journal} {\bibinfo  {journal} {JCAP}\ }\textbf {\bibinfo {volume} {1804}},\
  \bibinfo {pages} {016} (\bibinfo {year} {2018})},\ \Eprint
  {http://arxiv.org/abs/1612.08270} {arXiv:1612.08270 [astro-ph.CO]}
  \BibitemShut {NoStop}%
\bibitem [{\citenamefont {Ballardini}\ \emph {et~al.}(2016)\citenamefont
  {Ballardini}, \citenamefont {Finelli}, \citenamefont {Fedeli},\ and\
  \citenamefont {Moscardini}}]{Ballardini:2016hpi}%
  \BibitemOpen
  \bibfield  {author} {\bibinfo {author} {\bibfnamefont {Mario}\ \bibnamefont
  {Ballardini}}, \bibinfo {author} {\bibfnamefont {Fabio}\ \bibnamefont
  {Finelli}}, \bibinfo {author} {\bibfnamefont {Cosimo}\ \bibnamefont
  {Fedeli}}, \ and\ \bibinfo {author} {\bibfnamefont {Lauro}\ \bibnamefont
  {Moscardini}},\ }\bibfield  {title} {\enquote {\bibinfo {title} {{Probing
  primordial features with future galaxy surveys}},}\ }\href {\doibase
  10.1088/1475-7516/2018/04/E01, 10.1088/1475-7516/2016/10/041} {\bibfield
  {journal} {\bibinfo  {journal} {JCAP}\ }\textbf {\bibinfo {volume} {1610}},\
  \bibinfo {pages} {041} (\bibinfo {year} {2016})},\ \bibinfo {note} {[Erratum:
  JCAP1804,no.04,E01(2018)]},\ \Eprint {http://arxiv.org/abs/1606.03747}
  {arXiv:1606.03747 [astro-ph.CO]} \BibitemShut {NoStop}%
\bibitem [{\citenamefont {Ballardini}\ \emph {et~al.}(2018)\citenamefont
  {Ballardini}, \citenamefont {Finelli}, \citenamefont {Maartens},\ and\
  \citenamefont {Moscardini}}]{Ballardini:2017qwq}%
  \BibitemOpen
  \bibfield  {author} {\bibinfo {author} {\bibfnamefont {Mario}\ \bibnamefont
  {Ballardini}}, \bibinfo {author} {\bibfnamefont {Fabio}\ \bibnamefont
  {Finelli}}, \bibinfo {author} {\bibfnamefont {Roy}\ \bibnamefont {Maartens}},
  \ and\ \bibinfo {author} {\bibfnamefont {Lauro}\ \bibnamefont {Moscardini}},\
  }\bibfield  {title} {\enquote {\bibinfo {title} {{Probing primordial features
  with next-generation photometric and radio surveys}},}\ }\href {\doibase
  10.1088/1475-7516/2018/04/044} {\bibfield  {journal} {\bibinfo  {journal}
  {JCAP}\ }\textbf {\bibinfo {volume} {1804}},\ \bibinfo {pages} {044}
  (\bibinfo {year} {2018})},\ \Eprint {http://arxiv.org/abs/1712.07425}
  {arXiv:1712.07425 [astro-ph.CO]} \BibitemShut {NoStop}%
\bibitem [{\citenamefont {Palma}\ \emph {et~al.}(2018)\citenamefont {Palma},
  \citenamefont {Sapone},\ and\ \citenamefont {Sypsas}}]{Palma:2017wxu}%
  \BibitemOpen
  \bibfield  {author} {\bibinfo {author} {\bibfnamefont {Gonzalo~A.}\
  \bibnamefont {Palma}}, \bibinfo {author} {\bibfnamefont {Domenico}\
  \bibnamefont {Sapone}}, \ and\ \bibinfo {author} {\bibfnamefont {Spyros}\
  \bibnamefont {Sypsas}},\ }\bibfield  {title} {\enquote {\bibinfo {title}
  {{Constraints on inflation with LSS surveys: features in the primordial power
  spectrum}},}\ }\href {\doibase 10.1088/1475-7516/2018/06/004} {\bibfield
  {journal} {\bibinfo  {journal} {JCAP}\ }\textbf {\bibinfo {volume} {1806}},\
  \bibinfo {pages} {004} (\bibinfo {year} {2018})},\ \Eprint
  {http://arxiv.org/abs/1710.02570} {arXiv:1710.02570 [astro-ph.CO]}
  \BibitemShut {NoStop}%
\bibitem [{\citenamefont {Gong}\ \emph {et~al.}(2017)\citenamefont {Gong},
  \citenamefont {Palma},\ and\ \citenamefont {Sypsas}}]{Gong:2017vve}%
  \BibitemOpen
  \bibfield  {author} {\bibinfo {author} {\bibfnamefont {Jinn-Ouk}\
  \bibnamefont {Gong}}, \bibinfo {author} {\bibfnamefont {Gonzalo~A.}\
  \bibnamefont {Palma}}, \ and\ \bibinfo {author} {\bibfnamefont {Spyros}\
  \bibnamefont {Sypsas}},\ }\bibfield  {title} {\enquote {\bibinfo {title}
  {{Shapes and features of the primordial bispectrum}},}\ }\href {\doibase
  10.1088/1475-7516/2017/05/016} {\bibfield  {journal} {\bibinfo  {journal}
  {JCAP}\ }\textbf {\bibinfo {volume} {1705}},\ \bibinfo {pages} {016}
  (\bibinfo {year} {2017})},\ \Eprint {http://arxiv.org/abs/1702.08756}
  {arXiv:1702.08756 [astro-ph.CO]} \BibitemShut {NoStop}%
\bibitem [{\citenamefont {Cheung}\ \emph {et~al.}(2008)\citenamefont {Cheung},
  \citenamefont {Creminelli}, \citenamefont {Fitzpatrick}, \citenamefont
  {Kaplan},\ and\ \citenamefont {Senatore}}]{Cheung:2007st}%
  \BibitemOpen
  \bibfield  {author} {\bibinfo {author} {\bibfnamefont {Clifford}\
  \bibnamefont {Cheung}}, \bibinfo {author} {\bibfnamefont {Paolo}\
  \bibnamefont {Creminelli}}, \bibinfo {author} {\bibfnamefont {A.~Liam}\
  \bibnamefont {Fitzpatrick}}, \bibinfo {author} {\bibfnamefont {Jared}\
  \bibnamefont {Kaplan}}, \ and\ \bibinfo {author} {\bibfnamefont {Leonardo}\
  \bibnamefont {Senatore}},\ }\bibfield  {title} {\enquote {\bibinfo {title}
  {{The Effective Field Theory of Inflation}},}\ }\href {\doibase
  10.1088/1126-6708/2008/03/014} {\bibfield  {journal} {\bibinfo  {journal}
  {JHEP}\ }\textbf {\bibinfo {volume} {03}},\ \bibinfo {pages} {014} (\bibinfo
  {year} {2008})},\ \Eprint {http://arxiv.org/abs/0709.0293} {arXiv:0709.0293
  [hep-th]} \BibitemShut {NoStop}%
\bibitem [{\citenamefont {Adshead}\ \emph {et~al.}(2011)\citenamefont
  {Adshead}, \citenamefont {Hu}, \citenamefont {Dvorkin},\ and\ \citenamefont
  {Peiris}}]{Adshead:2011bw}%
  \BibitemOpen
  \bibfield  {author} {\bibinfo {author} {\bibfnamefont {Peter}\ \bibnamefont
  {Adshead}}, \bibinfo {author} {\bibfnamefont {Wayne}\ \bibnamefont {Hu}},
  \bibinfo {author} {\bibfnamefont {Cora}\ \bibnamefont {Dvorkin}}, \ and\
  \bibinfo {author} {\bibfnamefont {Hiranya~V.}\ \bibnamefont {Peiris}},\
  }\bibfield  {title} {\enquote {\bibinfo {title} {{Fast Computation of
  Bispectrum Features with Generalized Slow Roll}},}\ }\href {\doibase
  10.1103/PhysRevD.84.043519} {\bibfield  {journal} {\bibinfo  {journal} {Phys.
  Rev.}\ }\textbf {\bibinfo {volume} {D84}},\ \bibinfo {pages} {043519}
  (\bibinfo {year} {2011})},\ \Eprint {http://arxiv.org/abs/1102.3435}
  {arXiv:1102.3435 [astro-ph.CO]} \BibitemShut {NoStop}%
\bibitem [{\citenamefont {Maldacena}(2003)}]{Maldacena:2002vr}%
  \BibitemOpen
  \bibfield  {author} {\bibinfo {author} {\bibfnamefont {Juan~Martin}\
  \bibnamefont {Maldacena}},\ }\bibfield  {title} {\enquote {\bibinfo {title}
  {{Non-Gaussian features of primordial fluctuations in single field
  inflationary models}},}\ }\href {\doibase 10.1088/1126-6708/2003/05/013}
  {\bibfield  {journal} {\bibinfo  {journal} {JHEP}\ }\textbf {\bibinfo
  {volume} {05}},\ \bibinfo {pages} {013} (\bibinfo {year} {2003})},\ \Eprint
  {http://arxiv.org/abs/astro-ph/0210603} {arXiv:astro-ph/0210603 [astro-ph]}
  \BibitemShut {NoStop}%
\bibitem [{\citenamefont {Wang}(2014)}]{Wang:2013eqj}%
  \BibitemOpen
  \bibfield  {author} {\bibinfo {author} {\bibfnamefont {Yi}~\bibnamefont
  {Wang}},\ }\bibfield  {title} {\enquote {\bibinfo {title} {{Inflation, Cosmic
  Perturbations and Non-Gaussianities}},}\ }\href {\doibase
  10.1088/0253-6102/62/1/19} {\bibfield  {journal} {\bibinfo  {journal}
  {Commun. Theor. Phys.}\ }\textbf {\bibinfo {volume} {62}},\ \bibinfo {pages}
  {109--166} (\bibinfo {year} {2014})},\ \Eprint
  {http://arxiv.org/abs/1303.1523} {arXiv:1303.1523 [hep-th]} \BibitemShut
  {NoStop}%
\bibitem [{\citenamefont {Arroja}\ and\ \citenamefont
  {Koyama}(2008)}]{Arroja:2008ga}%
  \BibitemOpen
  \bibfield  {author} {\bibinfo {author} {\bibfnamefont {Frederico}\
  \bibnamefont {Arroja}}\ and\ \bibinfo {author} {\bibfnamefont {Kazuya}\
  \bibnamefont {Koyama}},\ }\bibfield  {title} {\enquote {\bibinfo {title}
  {{Non-gaussianity from the trispectrum in general single field inflation}},}\
  }\href {\doibase 10.1103/PhysRevD.77.083517} {\bibfield  {journal} {\bibinfo
  {journal} {Phys. Rev.}\ }\textbf {\bibinfo {volume} {D77}},\ \bibinfo {pages}
  {083517} (\bibinfo {year} {2008})},\ \Eprint {http://arxiv.org/abs/0802.1167}
  {arXiv:0802.1167 [hep-th]} \BibitemShut {NoStop}%
\bibitem [{\citenamefont {Chen}\ \emph {et~al.}(2009)\citenamefont {Chen},
  \citenamefont {Hu}, \citenamefont {Huang}, \citenamefont {Shiu},\ and\
  \citenamefont {Wang}}]{Chen:2009bc}%
  \BibitemOpen
  \bibfield  {author} {\bibinfo {author} {\bibfnamefont {Xingang}\ \bibnamefont
  {Chen}}, \bibinfo {author} {\bibfnamefont {Bin}\ \bibnamefont {Hu}}, \bibinfo
  {author} {\bibfnamefont {Min-xin}\ \bibnamefont {Huang}}, \bibinfo {author}
  {\bibfnamefont {Gary}\ \bibnamefont {Shiu}}, \ and\ \bibinfo {author}
  {\bibfnamefont {Yi}~\bibnamefont {Wang}},\ }\bibfield  {title} {\enquote
  {\bibinfo {title} {{Large Primordial Trispectra in General Single Field
  Inflation}},}\ }\href {\doibase 10.1088/1475-7516/2009/08/008} {\bibfield
  {journal} {\bibinfo  {journal} {JCAP}\ }\textbf {\bibinfo {volume} {0908}},\
  \bibinfo {pages} {008} (\bibinfo {year} {2009})},\ \Eprint
  {http://arxiv.org/abs/0905.3494} {arXiv:0905.3494 [astro-ph.CO]} \BibitemShut
  {NoStop}%
\bibitem [{\citenamefont {Smith}\ \emph {et~al.}(2015)\citenamefont {Smith},
  \citenamefont {Senatore},\ and\ \citenamefont {Zaldarriaga}}]{Smith:2015uia}%
  \BibitemOpen
  \bibfield  {author} {\bibinfo {author} {\bibfnamefont {Kendrick~M.}\
  \bibnamefont {Smith}}, \bibinfo {author} {\bibfnamefont {Leonardo}\
  \bibnamefont {Senatore}}, \ and\ \bibinfo {author} {\bibfnamefont {Matias}\
  \bibnamefont {Zaldarriaga}},\ }\bibfield  {title} {\enquote {\bibinfo {title}
  {{Optimal analysis of the CMB trispectrum}},}\ }\href@noop {} {\  (\bibinfo
  {year} {2015})},\ \Eprint {http://arxiv.org/abs/1502.00635} {arXiv:1502.00635
  [astro-ph.CO]} \BibitemShut {NoStop}%
\end{thebibliography}%

\end{document}